\documentclass[12pt]{article}
  \hoffset = - 1 truecm
\def\beq{\begin{equation}}   \def\eeq{\end{equation}}
\def\bea{\begin{eqnarray}}  \def\eea{\end{eqnarray}} 
\def\noi{\noindent} \def\beeq{\begin{eqnarray}}
\def\eeeq{\end{eqnarray}}
\def\lsim{\raise0.3ex\hbox{$<$\kern-0.75em\raise-1.1ex\hbox{$\sim$}}}
\def\gsim{\raise0.3ex\hbox{$>$\kern-0.75em\raise-1.1ex\hbox{$\sim$}}}

\newcommand\mysection{\setcounter{equation}{0}\section}

\newcounter{hran}

\begin{document}
\begin{center} \vbox to 1 truecm {} {\large \bf
Decoherence, Irreversibility and the Selection by
Decoherence of Quantum States with Definite Probabilities} \vskip 1 
truecm {\bf Roland Omn\`es} \vskip 3 truemm

{\it Laboratoire de Physique Th\'eorique\footnote{Unit\'e Mixte de
Recherche CNRS - UMR N$^{\circ}$ 8627},\\ Universit\'e Paris XI,
B\^atiment 210, 91405 Orsay Cedex, France} \end{center} \vskip 2 truecm

\begin{abstract}
      The problem investigated in this paper is einselection, i. e. the
selection of mutually exclusive quantum states with definite
probabilities through decoherence. Its study is based on a theory of
decoherence resulting from the projection method in the quantum theory
of irreversible processes, which is general enough for giving
reliable predictions. This approach leads to a definition (or
redefinition) of the coupling with the environment involving only
fluctuations. The range of application of perturbation calculus is then
wide, resulting in a rather general master equation. \par

Two distinct cases of decoherence are then found: (i) A ``degenerate''
case (already encountered with solvable models) where decoherence
amounts essentially to approximate diagonalization; (ii) A general
case where the einselected states are essentially classical. They are
mixed states. Their density operators are proportional to
microlocal projection operators (or ``quasi projectors'') which were
previously introduced in the quantum expression of classical properties.\par

It is found at various places that the main limitation in our
understanding of decoherence is the lack of a systematic method for 
constructing
collective observables.
\end{abstract} \vskip 1
truecm

\noi PACS Numbers : 03.65 Th, 03.65.Db, 05.40-a \par \vskip 5 truemm

\noi To be published in Phys. Rev. A \par
\vskip 1 truecm
\noi LPT Orsay 02-02 \par
\noi January 2002

\newpage \baselineskip=24 pt \pagestyle{plain}  \mysection{Introduction}
\hspace*{\parindent}
The discovery of decoherence has already much improved our
understanding of quantum mechanics. The effect has now been observed
experimentally \cite{1r}. Many of its consequences have been obtained
theoretically, but its foundation, the range of its validity and its
full meaning are still rather obscure. This is due most
probably to the fact that it deals with deep aspects of physics, not 
yet fully investigated.\par

The intuitive idea of decoherence is rather clear \cite{2r}. The wave
function of a macroscopic system depends on a very large number of
variables and its local phase is very sensitive to boundary conditions,
couplings and initial conditions. Any phase coherence between different
components of the wave function is therefore exposed to destruction,
after which macroscopic interferences disappear. It is unfortunately 
very difficult to build up a satisfactory theory on this intuition,
because a knowledge of phases remains out of reach for the $N$-body
methods at our disposal. \\

\noi {\large \it Some questions about decoherence}\\

      The problems of decoherence are most often stated after making a 
few simple
and pragmatic assumptions. One assumes particularly that
a few collective (or relevant) observables can describe the
main features of a (generally macroscopic) system, and they
are known {\it a priori}. The system is then split formally into
two subsystems: a ``collective'' one (which is associated with the
relevant observables) and an environment, which can be external or
internal. Each of these two abstract subsystems has its own Hilbert
space, ${\cal H}_c$ or ${\cal H}_e$ and the Hilbert space of the 
whole system is the
product ${\cal H} = {\cal H}_c \otimes {\cal H}_e$. The Hamiltonian 
is accordingly split into three
parts, one for each subsystem and one for their coupling:
\begin{equation}
\label{1.1e}
H = H_c \otimes I_e + I_c \otimes H_e + H_1 \ .
\end{equation}

\noi Observers are supposed to have only a direct knowledge of the
collective subsystem. Its properties are expressed by a reduced density
operator $\rho_r$, which is obtained from the full density operator 
$\rho$ through
a partial trace on the environment
\begin{equation}
\label{1.2e}
\rho_r = Tr_e\rho \ .
\end{equation}

\noi The time evolution of $\rho_r$ exhibits the various aspects of 
decoherence. It
has been investigated mostly by means of more or less exactly
solvable models. Two models were particularly important because they
were rather close to reality, at least in specific circumstances. In one
of them the environment is replaced by a collection of harmonic
oscillators [3-8]. Another model represents decoherence as an
accumulation of scattering phase shifts when particles from an external
atmosphere collide with a macroscopic object \cite{9r}. Much of what is known
about decoherence was learned from these models and some of their
variants \cite{10r}.\par

The conclusions have been accurately summarized by Zurek. He
distinguishes three different physical effects resulting from
decoherence\cite{10r}: There is first a destruction of macroscopic
interferences, then some privileged state vectors become selected as
alternative physical events and finally these states evolve
classically. The privileged states are also called pointer states in
analogy with the position of a pointer on a dial in a measuring
apparatus \cite{11r}. Most models predict that these states exist and are
orthogonal so that they define a Hilbert space basis in which the
reduced density operator becomes approximately diagonal after a short
while. The existence of this basis is essential since it defines a
unique set of alternative events with well-defined probabilities. The
name of ``einselection'' has been coined by Zurek for the mechanism
selecting this basis.\par

These results have so far-reaching consequences for the interpretation
of quantum mechanics and other applications such as quantum computing
that one must assess their exact degree of generality. How much of them
is specific to the models that were used and how much is universal~?
This question raises several problems:\\

\noi $\bullet$ 1. A basic preliminary problem is concerned with the meaning of
collective observables. When an actual physical system is given in
practice, it is a rather straightforward matter to guess what coordinates
describe most conveniently its macroscopic dynamics (the choice
of these ``generalized coordinates'' goes back to Lagrange). But
the question of defining correctly the collective observables for an
arbitrary quantum state of the whole system, i. e. to select what is
collective and what can be considered as an environment, is much
deeper. It will be seen again and again in this paper that it
represents the real limit of our understanding.\\

\noi $\bullet$ 2. One may be puzzled by the fact that explicit models yield
einselection somewhat too easily. This is because most of them rely on
a coupling satisfying the commutation property
\begin{equation}
\label{1.3e}
\left [ H_1, X \otimes I_e \right ] = 0  \ ,
\end{equation}

\noi between the coupling Hamiltonian $H_1$ and a collective coordinate
observable $X$ (which may be supposed unique for simplicity). It is
then found that $\rho_r$ becomes approximately diagonal in the basis $|x>$  of
eigenvectors of $X$. It is clear however than Eq. (\ref{1.3e}) is very
restrictive, at least from a mathematical standpoint, and one cannot
assume it to hold universally. What happens then when this condition is
not satisfied~? Is there still some sort of diagonalization~? If so,
along which basis~? To answer this problem will be the main task of this
paper.\\

\noi $\bullet$ 3. Problem 2 is made somewhat tricky because there 
exists a very large class
of systems with collective coordinates for which condition (\ref{1.3e}) holds.
They are mechanical non-relativistic systems (excluding
macroscopic electromagnetic effects), described by the position
coordinates of hydrodynamics [12-15]. These systems, which are
exceptional from a mathematical standpoint, may very well turn to be
universal in measurement theory since a measuring device involves
practically always some mechanical parts entangled with the rest of the
apparatus. As will be shown in Section 7, the property (\ref{1.3e}) results
from the form of kinetic energies and two-body potentials in
non-relativistic physics. This remarkable feature ``explains'' why
classical mechanics can be formulated in ordinary three-dimensional
space although the wave functions are defined on a configuration space
\cite{16r}.

So frequent an occurrence of a very special case may be puzzling from
an intuitive standpoint. It raises a conflict between what
we consider as general, either when speaking of the physical world or of the
mathematics of the theory. This possible source of confusion will be 
avoided here
by referring explicitly to ``mathematical generality'' when a ``general
property'' or a ``general case'' will be mentioned, except when otherwise
explicitly stated.\\

\noi $\bullet$ 4. One might be lured by models into premature 
conclusions and a sufficiently
wide-ranging theory of decoherence is necessary for assessing
general properties. The construction of such a theory is the fourth
problem to be considered here.\\

\noi $\bullet$ 5. Finally, one must consider the attractive approach 
to einselection by
Zurek \cite{10r,17r}. Einselected states are supposed to be the most
predictable (or robust) carriers of information. Given a collective
state $\Psi$ (which may be pure or not) and the corresponding initial
density operator $\rho_{\Psi} (0) = |\Psi> < \Psi|$, one considers 
the time-dependent reduced
density operator $\rho_{\psi} (t)$. Its ability to preserve an 
information content
is characterized by some relevant functional of $\Psi$, which may be 
minus the von
Neumann entropy or more conveniently the measure of purity
\begin{equation}
\label{1.4e}
c_{\Psi}(t) = Tr \rho_{\Psi}^2 (t)  \ .
\end{equation}

This quantity is then used to construct a ``predictability sieve''
distinguishing among the states: The largest the quantity (\ref{1.4e}) is, the
more predictable the state $\Psi$ is supposed to be. Model examples suggest
that einselected states do exist and are rather insensitive to a change
in the coupling or a redefinition of the environment. A fifth problem
consists in evaluating this conjecture in a wider framework.\\

\noi {\large \it The present results}\\

These five problems will not of course be solved here completely, but
some definite or suggestive answers will be obtained. The most precise
results are concerned with einselection and diagonalization, their
meaning and their relation. As a preliminary, one needs a sufficiently
wide-ranging theory of decoherence, as stated in Problem 4. The theory
to be used here does not claim to be new. It relies on the
familiar idea that decoherence is a special kind of irreversible
process. This means that one may expect that the most general theory of
decoherence presently at our disposal would derive from the existing
quantum theories of irreversible processes. Moreover, the most convenient
such theory is the so-called projection method [18-21]. Its
main features are recalled in Section 2 and it is applied to
decoherence in Section 3. Although this method was previously
introduced elsewhere \cite{15r,22r,23r}, some improvements will be required
before applying it for the present purpose. These developments are
mostly given in the Appendices and they may be useful for
using the method in other problems.\par

A very important remark concerning this approach is the possibility of
making a definite choice for the coupling Hamiltonian $H_1$ from which one
can easily derive a master equation for $\rho_r$. The point is that, 
given {\it a
priori} a coupling Hamiltonian, one can construct another
(time-dependent) one consisting only of fluctuations. Standard
perturbation methods can then be applied confidently in most cases.
Although this procedure is familiar near equilibrium (it is used
for instance in the derivation of the fluctuation-dissipation theorem
\cite{24r}), it works also far from equilibrium as will be shown in Section 4,
providing a strong handle on decoherence. \par

The master equation one obtains in this way is probably the most
general one that is accessible with present techniques and therefore
the most appropriate one for investigating einselection, as done
in Section 5. One thus finds that, contrary to current
expectations, two different cases of einselection exist. The first one
was encountered with solvable models and it is well known,
although it is far from being general in a mathematical sense (in the
case of an arbitrary coupling). It must be considered on the contrary
as a degenerate case for the following reason: If $n$ is the number of
$X$-observables, decoherence is controlled in the general case by a
differential Laplacian-like operator in the $2n$-dimensional phase space.
In the simplest case $n = 1$, the decoherence ``Laplacian'' is associated
as usual with a quadratic form (like the two-dimensional Laplacian 
$\partial^2/\partial x^2 + \partial^2/\partial p^2$ is
associated after a Fourier transform $(x, p) \to (\xi , \eta )$ to 
the quadratic
form $(\xi^2 + \eta^2$). The decoherence Laplacian is degenerate when 
it acts on only
one variable (for instance the coordinate $x$ and not the associated
momentum $p$) so that the corresponding quadratic form is degenerate
(having a zero eigenvalue). In the degenerate case, to which the
usual models belong, decoherence essentially amounts to a
diagonalization (in the basis einselected by degeneracy).\par

The non-degenerate case is investigated in Section 6. The results are
not those expected from Zurek's predictability sieve, at least as
far as I understand it. There are generally two distinct times for
decoherence. Typically, in conditions when these times are very
different, decoherence selects a basis of privileged states in which $\rho_r$
begins to become diagonal, but then these ``einselected'' states do not
preserve their probabilities and they begin to share them with neighboring
states. When the two decoherence times are similar, not much remains 
apparently of the idea of einselection. The final outcome of 
decoherence
is rather a tendency towards uniformity where $\rho_r$  becomes as close to
the identify matrix as its finite trace can allow (more precisely, it
corresponds to a uniform Wigner function over a rather large
region of phase space). Nevertheless, macroscopic interferences are
still destroyed and classical behavior may follow.\par

The question of the different time scales is discussed in section 7 and
a strong connection is found with the special properties of
hydrodynamical variables. The relation between decoherence and
dissipation coefficients plays an important role in this discussion.
Problem 3 becomes then central because decoherence depends most often
in practice on the fact that the space coordinates of a
non-relativistic piece of matter satisfy the condition (\ref{1.4e}) implying
degeneracy. One is thus led again to Gell-Mann and Hartle's
ideas concerning the link between coarse graining and the existence 
of a diagonalization basis \cite{12r,13r}.\par

Finally, the occurrence of classical dynamics after decoherence is
considered in Section 8 in both the degenerate and the non-degenerate
cases. In the non-degenerate case, although one can still speak of
einselected states, they are far from being pure states. They are mixed
states whose density operator at a definite time is proportional to a
``microlocal projection operator'', which is known in
mathematics as the best expression of a classical property involving
position and momentum together [25-27]. Finally, the conclusion goes
back to the strong connection between the origin of classicality
and Problem 1, i. e. the construction of collective coordinates.
Some proposals for further research are indicated.\par

Appendix A gives a derivation of decoherence theory from the projection
method in irreversible processes. Appendix B is concerned with the
relation between decoherence and dissipation coefficients. Appendix C
shows how to include the insightful scattering model of decoherence by
Joos and Zeh in the framework of the present theory.

\mysection{A quantum theory of irreversible processes}
\hspace*{\parindent}
One needs a sufficiently wide-ranging theory of decoherence for
asserting its general features. Since the loss of phase coherence through
decoherence produces disorder, typical of an irreversible
process, the most general relevant theory is the
projection method, which is now briefly reviewed [18-21].\par

One considers a system with many degrees of freedom, whose density
operator evolves according to the von Neumann-Schr\"odinger equation,
\beq \label{2.1e} \dot{\rho} = - {i \over \hbar} \ [H, \rho ]\ . \eeq

Some observables are supposed to be particularly relevant for a given
problem and one wants only to know the time evolution of their mean
values. These relevant observables are denoted by
$A^j$. Their set may be finite or not, countable or not. One usually
includes the identity operator $I$ in the set and also the conserved
quantities such as the total energy $H$, although this recipe is not
imperative. The ``exact'' average values of these
observables are
\beq \label{2.2e} a^j(t) = Tr\left ( A^j \rho (t) \right ) \ . \eeq

The first step of the method consists in introducing a time-dependent
test density operator $\rho_0 (t)$ satisfying the following two conditions: (i)
It gives the exact average values of the relevant observables:
\beq \label{2.3e} Tr \left ( A^j \rho_0 (t) \right ) = Tr \left ( A^j 
\rho (t) \right ) \equiv a^j (t) \ . \eeq

\noi (ii) Its information content is minimal (which means that it does not
provide anything else than the quantities $\{ a^j(t)\})$. It can therefore be
written as
\beq \label{2.4e} \rho_0(t) = \exp \left ( - \lambda_j(t) A^j \right )\eeq

\noi where the numbers $\{ \lambda_j\}$ are Lagrange parameters and 
the Einstein
summation convention over repeated indices has been used. Since the
identity $I$ belongs to the set $\{ A^j\}$, the trace of $\rho_0$ is 
normalized. \par

One will use the name ``density'' in the present paper for an operator
with a finite trace (also called a trace-class operator). It is
neither supposed to have a unit trace nor to be necessarily positive.
One defines a set of (time-dependent) densities
\beq \label{2.5e} s_j = \partial \rho_0 / \partial a' \ .  \eeq

\noi They satisfy the important orthogonality properties
\beq \label{2.6e} Tr\left ( s_iA^j\right ) = \delta_i ^j \ , \eeq

\noi amounting essentially to $\partial a^j/\partial a^i = 
\delta_i^j$ in view of Eq. (\ref{2.3e}).

The theory makes use of ``superoperators'', which act linearly on a
density to yield another density. For instance, Eq. (\ref{2.1e}) can be written
conventionally as
\beq \label{2.7e} \dot{\rho} = L \rho \ ,\eeq

\noi where $L$ is the Liouville superoperator. Another important
superoperator is defined by
\beq \label{2.8e} P = s_j \otimes A^j \ ,  \eeq

\noi which means that when acting on a density $\mu$, it gives (with the
summation convention)
\beq \label{2.9e}  P\mu = s_j \cdot Tr\left ( A^j \mu \right ) \ .\eeq

\noi It is a projection in so far as it satisfies the simple equation
\beq \label{2.10e} P^2 = P \ , \eeq

\noi resulting from the orthogonality properties (\ref{2.6e}).\par

One can define a ``relevant'' (time-dependent) density operator $\rho_1$ by
\beq \label{2.11e} \rho_1 = P \rho \ . \eeq

\noi It generates the exact average values $\{ a(t) \}$ since
\beq \label{2.12e} Tr \left ( A^j \rho_1 \right ) = Tr \left ( A^j P \rho \right ) = Tr \left ( A^j s_i \right ) Tr \left ( A^i \rho \right 
) = \delta_i^j a^i = a^j \ .\eeq

\noi (As a matter of fact, it coincides with the test density 
operator $\rho_0$). \par

Denoting by ${\cal I}$ the identity superoperator, one also 
introduces ${\cal Q} = {\cal I}  -
P$, which satisfies the projection property ${\cal Q}^2 = {\cal Q} $ 
in view of Eq. (\ref{2.10e}),
as well as the orthogonality properties ${\cal Q}P = P{\cal Q} = 0$. 
One defines then
another density $\rho_2 = {\cal Q} \rho$ (so that $\rho = \rho_1 + 
\rho_2$ ). Applying the two projections $P$ and ${\cal Q}$ to
the evolution equation (\ref{2.7e}) and taking into account the time
dependence of these projections, one obtains two coupled evolution
equations
\beq \label{2.13e} \dot{\rho}_1 = PLP\rho_1 + \dot{P}P\rho_1 + 
PL{\cal Q} \rho_2 + \dot{P} {\cal Q} \rho_2 \ ,        \eeq
\beq \label{2.14e} \dot{\rho}_2 = {\cal Q}L{\cal Q}\rho_2 + 
\dot{P}{\cal Q}\rho_2 + {\cal Q}LP \rho_1 - \dot{P} P \rho_1 \ .\eeq

\noi A last step would be to eliminate $\rho_2$ to obtain a master
equation for $\rho_1$  but it will be left for the special case of decoherence.

\mysection{The case of decoherence} \hspace*{\parindent}
The previous theory can be now applied to decoherence and some preliminary
considerations will make the task clearer. A first problem is to choose
the relevant observables. If one thinks of
macroscopic interferences, it is clear that they can involve many
different macroscopic observables so that every collective observable
is relevant. When the collective observables describe a measuring
apparatus, the measured microscopic observable is also relevant
although not collective. \par

The environment can be defined by its observables, which commute with
the collective ones. The resulting splitting of the system into a
collective one and an environment is generally time dependent since for
instance every new bubble in a bubble chamber or every new spark in a
spark chamber generates new collective observables. Decoherence is such
a rapid process however that the collective environment splitting can
often be considered as fixed during the very short time of an
individual decoherence process, justifying the expression (\ref{1.1e}) for the
Hamiltonian. The coupling $H_1$  in this equation is responsible for the
interactions between the collective system and the environment,
including dissipation and decoherence. \par

The set of relevant observables is completed by the identity operator $I$
(insuring normalization) and the environment Hamiltonian $H_e$, or more
properly $I_c \otimes H_e$. The total Hamiltonian $H$ might have been 
used as relevant in
place of $H_e$  but this choice would have been inconvenient, as will be
seen later). \\

Introducing an arbitrary orthonormal basis $|k>$ in the collective Hilbert
space, the set $\{|k><k'|\}$ provides a linear basis for the 
collective operators.
A look at the calculations in Section 2 shows that they nowhere use the
fact that the $\{A^j\}$  are hermitian so that one may use the set of operators
$\{|k><k'|\}$ as relevant ``observables''. Alternatively, one might 
use the set of
hermitian operators
$$\{(1/2)(|k> \pm |k'>)(<k|\pm <k'|);(1/2)(|k>\pm i|k'>)(<k|\mp i<k'|)\}$$

\noi as relevant with identical results. Anyway, the set of relevant
observables for a theory of decoherence will be chosen as
\beq \label{3.1e} \left \{ A^{kk'} = |k><k'| \otimes I_e , A^1 = I, 
A^e = I_c \otimes H_e \right \} \ . \eeq

\noi Since none of them connects the collective and the environment 
Hilbert spaces, the test density operator (\ref{2.4e}) is a tensor 
product
\beq \label{3.2e} \rho_0 = \rho_c \otimes \rho_e \ . \eeq

\noi Applying Eq. (\ref{2.3e}) to the operators $A^{kk'}$, one finds that
\beq \label{3.3e}
<k'|\rho_c | k > = Tr \left ( A^{kk'} \rho_0 \right ) \equiv Tr \left 
( A^{kk'} \rho \right ) = <k'|tr \rho |k> = <k'|\rho_r |k>
\eeq

\noi so that the collective test density $\rho_e$  is identical with the
conventional reduced density (\ref{1.2e}). The second equality results
from Eq. (\ref{2.1e}). A convention for traces that will be used everywhere has
also been introduced, the notation $Tr$ standing for a trace on the full
Hilbert space and $tr$ for a trace on the environment. \par

According to Eq. (\ref{2.4e}), the environment test density $\rho_e$ 
is given by
\beq \label{3.4e}
\rho_e = \exp \left ( - \beta H_e - \alpha \right ) \ ,
\eeq

\noi where $\alpha$ is a Lagrange parameter insuring normalization 
and $\beta$ insures
that the energy $H_e$ of the environment has its true average value $E$. This
density is the same as if the environment were in thermal equilibrium
but it should be stressed that it is only an auxiliary mathematical
quantity providing a correct (time-dependent) value for $E$ with no
assumption about equilibrium. \par

In Appendix A, the auxiliary densities $s_j$ are
obtained from Eq. (\ref{2.5e}). Denoting respectively by $s_1$ and 
$s_2$ the densities
associated with $I$ and $H_e$, one gets
\bea
\label{3.5e}
&&s_{kk'} = |k'><k|\otimes \rho_e \ , \\
&&s_e = \rho_c \otimes \rho_e \left ( H_e - E\right ) \Delta^{-2} \ , \\
\label{3.6e}
&&s_1 = - E \rho_c \otimes \rho_e \left ( H_e - E \right ) \Delta^{-2} \ ,
\label{3.7e}
\eea

\noi where $\Delta$ is the uncertainty in energy
\beq \label{3.8e}
\Delta^2 = tr \left ( H_e^2 \rho_e \right ) - E^2 \ .
\eeq

When acting on an arbitrary density $\mu$, the projection $P$ is 
given according to Eqs. (\ref{2.9e}) and (\ref{2.5e}-\ref{2.7e}) by
\beq \label{3.9e}
P\mu = tr \mu \otimes \rho_e + \left ( \rho_c \otimes \left \{ \rho_e 
( H_e - E) \Delta^{-2} \right \} \right ) \cdot \left ( TrH_e\mu - 
ETr \mu \right ) \ ,
  \eeq

\noi from which the relations $P^2 = P$ and $\rho_1 = \rho_0$ follow. \\

\noi {\large \it Specifying the coupling} \\

One may now introduce an important remark that will later justify the
use of perturbation theory. To begin with, one may notice some
arbitrariness in the splitting of the full Hamiltonian $H$ into three
different terms as in Eq. (\ref{1.1e}). A simple recipe for fixing them is to
impose that
\beq \label{3.10e}
tr H_1 \rho_e = 0 \ .
\eeq

\noi The meaning of this condition can be seen on the example of a
cylinder containing a gas. A collective coordinate is specified by
the position $x$ of a piston whereas the environment consists of the gas
and the matter of the piston itself. A straightforward definition of
$H_1$ could be the sum of the potential energies between the atoms in the
piston and the gas. This interaction is far from being weak, since
the confinement of a gas is not a weak effect, but a large part of it
consists of a collective energy since $trH_1\rho_e$  is a collective 
operator. One
can then change the definition of the different parts in $H$ by including
this operator in $H_c$ and removing it from $H_1$ or, more precisely, 
by introducing
\begin{eqnarray*}
&&H'_c = H_c + tr \left ( H_1 \rho_e \right ) \ , \\
&&H'_1 = H_1 - tr \left ( H_1 \rho_e \right ) \otimes I_e \ .
\end{eqnarray*}

\noi The quantity $trH_1 \rho_e$ represents in this example the effect of the
gas pressure on the piston. The new expression of $H_c$ is time-dependent
(like pressure) but the new expression of $H_1$ satisfies the condition
(\ref{3.10e}). It consists only of the pressure fluctuations resulting from
the collisions of the gas molecules with the piston.\par

  The idea of introducing a purely fluctuating coupling and to use perturbation
theory for computing its effects is familiar in quantum fluctuation
theory \cite{24r}. The fact that one can still use it far from equilibrium
when dealing with decoherence is due to the possibility of representing
everything collective by the test density. From there on, the condition
(\ref{3.10e}) will be assumed. \par

One may also understand at this point why the choice of $H_e$ as a 
relevant observable is more convenient
than the total Hamiltonian $H$, which is usually recommended \cite{21r}. This
is because the expression (\ref{3.2e}) for the test density implies the simple
rule (\ref{3.10e}) for the coupling, with the benefits just mentioned.
Everything would have been more obscure and would have implied much heavier
calculations if $H$ had been chosen as a relevant observable. \\

\noi {\large \it The evolution equations} \\

It is easy to write down explicitly the evolution equations
(\ref{2.13e}-\ref{2.14e}) for the case of decoherence. It is 
convenient to split Eq.
(\ref{2.13e}) for $\dot{\rho}_1$ into an equation for $\dot{\rho}_r$ 
and another for $\dot{\rho}_e$  (or for the time
evolution of the internal energy). This is done by taking respectively
the traces of Eq. (\ref{2.13e}) on the environment and the collective Hilbert
spaces. As shown in Appendix A, the results are:
\beq \label{3.11e}
\dot{\rho}_r = - {i \over \hbar} \left ( \left [ H_c, \rho_r \right ] 
+ tr \left [ H_1 , \rho_2 \right ] \right ) \ ,
\eeq
\beq \label{3.12e}
\dot{E} + {i \over \hbar} Tr \left ( H_e \left [ H_1 , \rho_1 + 
\rho_2 \right ] \right ) = 0 \ .
\eeq

\noi As for the second evolution equation (\ref{2.14e}), it becomes
\beq \label{3.14e}
\dot{\rho}_2 = - (i/\hbar ) \left [ H, \rho_1 + \rho_2 \right ] + 
(i/\hbar ) tr \left ( \left [ H_1 , \rho_2 \right ] \right ) \otimes 
\rho_e - \rho_r \otimes \dot{\rho}_e \ .
\eeq

\mysection{A master equation} \hspace*{\parindent} The most delicate
step in the projection method consists always in ``solving'' the second
evolution equation (\ref{2.14e}) for $\rho_2$ in terms of $\rho_1$ 
before inserting the
result into Eq. (\ref{2.13e}) \cite{21r}. This is much easier when perturbation
theory can be used. Perturbation calculus has been used already in the
present framework when $H_1$ is known {\it a priori} to be small, as often
happens in quantum optics \cite{22r,23r}. It should also presumably be valid
in many instances when condition (\ref{3.10e}) is applied and $H_1$ is a pure
fluctuation (although one must acknowledge that a purely fluctuating
coupling does not insure with certainty the validity of perturbation
calculus). Anyway, according to Appendix A, the evolution equations
(3.11-13) become at leading order in $H_1$~:
\beq \label{4.1e}
\dot{\rho}_r = - {i \over \hbar} \ \left ( \left [ H_c , \rho_r 
\right ] + tr \left [ H_1, \rho_2 \right ] \right ) \ ,
\eeq
\beq \label{4.2e}
\dot{\rho}_2 = - (i /\hbar) \left [ H_0 , \rho_2 \right ] -  (i 
/\hbar) \left [ H_1, \rho_2 \right ] \ .
\eeq

\noi In the second equation, $H_0$ denotes the uncoupled Hamiltonian
\beq \label{4.3e}
  H_0 = H_c \otimes I_e + I_c \otimes H_e
  \eeq

\noi and Eq. (\ref{4.1e}) is exact whereas Eq. (\ref{4.2e}) is valid 
at first order in perturbation theory.\par

      The second equation is easily solved after introducing the 
evolution operator
\beq \label{4.4e}
U(t) = \exp \left ( - iH_0t/\hbar \right ) \ .
\eeq

\noi Strictly speaking, $H_0$ is generally time dependent and the integrand
in Eq. (\ref{4.4e}) should be replaced by an integral of $H_0(t)$ on 
time. It is
difficult however to conceive of a case where this external time
dependence is not much slower than decoherence and the expression (\ref{4.4e})
is therefore most often valid as it stands. If not, the necessary
changes are so trivial that they need not be mentioned here. One thus
gets

\beq \label{4.5e}
\rho_2(t) = - (i/ \hbar ) \int_{-\infty}^t dt' \ U(t - t') \left [ 
H_1, \rho_1(t') \right ] U^{-1} (t - t') \ .
\eeq

\noi No effect of the initial value of $\rho_2$  (at time $- \infty$) 
has been included in
Eq. (\ref{4.5e}). This is justified when the environment is initially in
thermal equilibrium (since then $\rho_2 (-\infty ) = 0$). More 
generally however, it
may be expected that an initial lack of equilibrium does not influence
the decoherence effect, so that Eq. (\ref{4.5e}) is valid for our present
purpose. This point was checked in a special case by Paz,
Zurek and coworkers \cite{28r,29r}.\par

Inserting Eq. (\ref{4.5e}) into Eq. (\ref{4.1e}), one obtains the 
following ``master
equation'' for decoherence
\beq \label{4.6e}
\dot{\rho}_r = - {i \over \hbar} \left [ H_c , \rho_r \right ] - (1/ 
\hbar^2) \int_{-\infty}^t dt'\ tr \left [ H_1, U(t-t') \left [ H_1 , 
\rho_r (t') \otimes \rho_e(t') \right ] U^{-1} (t - t') \right ] \ .
\eeq

\noi The first term in the right-hand side represents the quantum evolution
of the reduced density operator under the action of the collective
Hamiltonian $H_c$. The second term is responsible for decoherence.\par

This equation is not new but it was derived previously either under the
assumption of a small coupling \cite{22r,23r}, or as a guess 
\cite{15r}. It will be
used here as a sufficiently general framework for a study of
einselection.\par

The wide range of this master equation is confirmed by its agreement
with previous models. This is easily shown when the environment is
represented by a collection of harmonic oscillators [3-8]. The
key experiment by Brune {\it et al.} showing the existence of
decoherence also belongs to the domain of Eq. (\ref{4.6e}) since $H_1$
is small in that case \cite{1r,30r}. In the case of the collision
model by Joos and Zeh the calculations are less trivial
and they are given in Appendix C as a non-trivial example of the 
master equation universality.

\mysection{Decoherence versus diagonalization} \hspace*{\parindent}
Models have been extremely useful for understanding
decoherence. When the collective subsystem is described by a few
position observables $X$, decoherence was found to diagonalize the
reduced density $\rho_r$, in the basis $|x>$ consisting of 
eigenvectors of $X$. The
question to be now considered is therefore: Does decoherence always
implies some sort of diagonalizationn ? Is there always a selection of
privileged ``pointer states'', or einselection as defined by Zurek 
\cite{10r}~?\par

One may first select the equation on which this question will be
investigated. The idea of diagonalization must be used with some care
because the reduced density operator never becomes completely diagonal
in view of the first term in Eq. (\ref{4.6e}) representing collective dynamics.
For a finite value of the difference $x - x'$ the matrix elements
\beq \label{5.1e}
\rho_r (x, x';t) = <x|\rho_r(t)|x'> \ ,
\eeq

\noi vanish exponentially with time, whereas microscopic values of $x - x'$ are
dominated by collective dynamics and they remain finite. This is why
there is decoherence on large scale while atomic physics remains
perfectly valid at small scale. The question of diagonalization is
therefore much clearer if one leaves aside the first term in Eq. (\ref{4.6e})
and consider ``pure decoherence'' as the behavior of a density operator
obeying the truncated equation
\beq \label{5.2e}
\dot{\rho}_r = - (1/\hbar^2) \int_{-\infty}^t dt' \ tr \left [H_1, 
U(t-t') \left [ H_1, \rho_r(t') \otimes \rho_e (t') \right ] U^{-1} 
(t - t') \right ] \equiv D \ .
\eeq

The main task will then consist in an analysis of the right-hand side 
of Eq. (\ref{5.2e})), which has been denoted by $D$. It will also be 
useful to introduce the notation
\beq \label{5.3e}
H_1^T = U(t-t')H_1U^{-1}(t-t') \ ,
\eeq

\noi so that one has
\beq \label{5.4e}
D = - (1/\hbar^2) \int_{-\infty}^t dt' \ tr \left [ H_1, \left [ 
H_1^T, U_c (t-t') \rho_r (t') U_c^{-1} (t-t') \otimes \rho_e(t') 
\right ] \right ]
\eeq

\noi (where $U_e^{-1}\rho_e U_e$ has been replaced by $\rho_e$ in 
view of Eq. (\ref{3.4e}). \\

\noi {\large \it Weyl symbols} \\

One will consider the case when there exists a set of $n$ collective
``position'' observables, altogether denoted by $X$. The quantity $D$
is itself a collective operator and it will be convenient to describe
it by means of a Weyl symbol \cite{31r,25r}, in analogy with the
description of $\rho_r$ by a Wigner function \cite{32r}. The standard
Weyl calculus can be slightly generalized to include ``operator-valued
symbols'' acting on the environment as follows:\par

Let $A$ denote an arbitrary operator in the full Hilbert space (such as
$H_1$ for instance). Introducing the basis $\{|x>\}$ in the collective Hilbert
space and an orthonormal basis $\{|n>\}$ in the environment Hilbert space,
the matrix elements of $A$ can be expressed through a partial Fourier
transform
\beq \label{5.5e}
<x, n|A|x',n'> = \int (2 \pi \hbar )^{-n} dp \overline{A}_{nn'}\left 
( {x + x' \over 2}, p\right ) \exp \left \{ ip (x' - x)\hbar \right 
\} \ .
\eeq

\noi Every quantity $\overline{A}_{nn'}(x,p)$ is a function of $(x, 
p)$ and the ordinary Weyl-symbol of
the matrix element $A_{nn'} = <n|A|n'>$, which is a collective 
operator. It will be
convenient to consider it as the $(n, n')$ matrix element of an
operator-valued symbol $\overline{A}(x,p)$, which is a function of 
$(x, p)$ and an operator
in the Hilbert space of the environment.\par

The symbol of the product $AB$ of two operators $A$ and $B$ can then be
expressed as a series in powers of $\hbar$ involving their symbols 
\cite{31r,25r}:
\beq \label{5.6e}
\overline{AB} = \overline{A}\cdot \overline{B} - (i \hbar/2) \left ( 
\overline{A}_p\overline{B}_x- \overline{A}_x \overline{B}_p \right ) 
- (\hbar^2 /24) \left ( \overline{A}_{p^2} \overline{B}_{x^2} + 
\overline{A}_{x^2} \overline{B}_{p^2} - 2 \overline{A}_{px} 
\overline{B}_{px} \right ) + O(\hbar^3)\ . \eeq

\noi The notation has been simplified by omitting the arguments $(x, p)$ of
the symbols and lower indices stand for derivatives (for instance, 
$\overline{A}_{xp}$
stands for $\partial^2 \overline{A}(x,p)/\partial x \partial p$). Eq. 
(\ref{5.6e}) is well known in Weyl's calculus when the
symbols are ordinary functions. It is easily extended to
operator-valued symbols by considering matrix elements and a unique new
rule must be added to the case of functions: the order
of the operators in the product $AB$ must be respected in the products of
symbols and their derivatives. The symbol of the reduced density 
operator $\rho_r$ is
the Wigner function, which will be denoted by $W(x, p)$. It is not
operator-valued and commutes with other symbols.\par

The only further formula one will need from the Weyl calculus is the
expression of a complete trace:
\beq \label{5.7e}
Tr A = \int dx dp (2 \pi \hbar)^{-n} tr \overline{A} (x, p) \ .
\eeq
\vskip 3 truemm

\noi {\large \it Calculation of the decoherence term D}\\

It will be convenient to consider from there on the case of a unique
coordinate $X$ $(n = 1)$ although the generalization to arbitrary 
values of $n$ is
trivial. Applying Eq. (\ref{5.6e}) to the double commutator in Eq. 
(\ref{5.4e}), one
obtains the symbol $\overline{D}$ of the decoherence term $D$ at 
order $\hbar^2$, as shown in
Appendix B:
\beq \label{5.8e}
\overline{D} = \int_{-\infty}^t dt'\left ( \partial / \partial x 
\left ( C^{xx} W_x^T + C^{xp} W_p^T \right ) + \partial /\partial p 
\left ( C^{px}W_x^T + C^{pp} W_p^T \right ) \right ) \ .
\eeq

\noi The function $W^T(x,p)$ is the symbol of the collective operator
\beq \label{5.9e}
U_c(t-t') \rho_r (t') U_c^{-1} (t-t') \ .
\eeq

\noi The various decoherence coefficients are given by
\bea \label{5.10e}
&&C^{xx} = {1 \over 2} tr \left \{ \left ( \overline{H}_{1p} 
\overline{H}_{1p}^T + \overline{H}_{1p}^T \overline{H}_{1p} \right ) 
\rho_e \right \} \ , \\
&&C^{xp} = - {1 \over 2} tr \left \{ \left ( \overline{H}_{1p} 
\overline{H}_{1x}^T + \overline{H}_{1x}^T \overline{H}_{1p} \right ) 
\rho_e \right \} \ ,\\
\label{5.11e}
   &&C^{px} = - {1 \over 2} tr \left \{ \left ( \overline{H}_{1x} 
\overline{H}_{1p}^T + \overline{H}_{1p}^T \overline{H}_{1x} \right ) 
\rho_e \right \} \ ,\\
\label{5.12e}
&&C^{pp} = {1 \over 2} tr \left \{ \left ( \overline{H}_{1x} 
\overline{H}_{1x}^T + \overline{H}_{1x}^T \overline{H}_{1x} \right ) 
\rho_e \right \} \ .
\label{5.13e}
\eea

  It is possible in principle to derive the main consequences of the
master equation for decoherence from these equations by using the
powerful methods of microlocal analysis \cite{25r}. It will be much
simpler however to rely on a few usual approximations. The first one
assumes that the coefficients (5.10-13) vary slowly with $(x, p)$ or,
more precisely, one neglects the collective evolution $U_c(t-t')$ in 
the factors
$U$ and $U^{-1}$ occurring in the expression (\ref{5.3e}) of $H_1^T$. 
The physical meaning of
this approximation is discussed in Appendix B, where the following
expression of $C^{pp}$ is obtained:
\beq \label{5.14e}
C^{pp} = \sum_{nn'} \overline{H}_{1xnn'} \ \overline{H}_{1xn'n} \exp \left ( i \omega_{nn'}(t - t') \right ) p_{nn'} \cosh \left ( \beta \hbar 
\omega_{nn'} /2 \right ) \ ,
\eeq

\noi where the states $|n>$ are the energy eigenstates of $H_e$ with 
eigenvalues $E_n$ and one
has written
\beq \label{5.15e}
\overline{H}_{1xnn'} = <n|\partial \overline{H}_1(x,p)/\partial 
x|n'>, \omega_{nn'} = \left ( E_n - E_{n'} \right ) / \hbar, p_{nn'} 
= \exp \left [ - \beta \left ( E_n + e_{n'}\right )/2 - \alpha \right 
] \ .
  \eeq

Eq. (\ref{5.14e}) suggests that the relevant frequencies 
$\omega_{nn'}$ in the sum are
contained in an interval $[-\Omega, \Omega ]$ characterizing the 
environment and
generally large as compared with the rate of collective dynamics ($\Omega$
is typically a Debye frequency for an internal environment). Hu, Paz
and Zhang have shown that the master equation is instantaneous (i. e.
involves no retardation) in the case an oscillator environment, when
the collective Hamiltonian also describes an oscillator \cite{7r}. This is
due to the linear character of the equations in that case \cite{33r}. The
resulting master equation has been solved explicitly by Ford and
O'Connell \cite{8r}. This situation is however exceptional and the neglect of
retardation is almost always an approximation. The question of its
justification is tricky and it would warrant a separate investigation.
When retardation effects are unimportant anyway, the integration on $t'$
in Eq. (\ref{5.8e}) is performed as if the integrand were a delta-function in
time. The time-delayed function $W^r$ is replaced by the ordinary Wigner
function $W$ and Eq. (\ref{5.8e}) becomes
\beq \label{5.16e}
\overline{D} = {\partial \over \partial x} \left ( g^{xx} W_x + 
g^{xp} W_p \right ) + {\partial \over \partial p} \left ( g^{p} W_x + 
g^{pp}W_p \right ) \ .
\eeq

\noi The new coefficients are given by
\beq \label{5.17e}
g^{ij} = \int_{-\infty}^t C^{ij}(t-t') dt'
\eeq

\noi (with indices $(i, j) = (x, p)$). Explicit expressions of these
coefficients are given in Appendix B, showing that the coefficients $g^{xx}$
and $g^{pp}$ are positive symmetric~: $g^{xp} = g^{px}$, and the quadratic form
\beq \label{5.18e}
g^{xx} \alpha^2 + 2g^{xp} \alpha \beta + g^{pp} \beta^2
\eeq

\noi is non-negative. One must then distinguish two significantly different
cases according to whether the form (\ref{5.18e}) is degenerate or not, i. e.
whether the determinant $g^{xx}g^{pp} - (g^{px})^2$  is zero or positive.\\

\noi {\large \it The degenerate case}\\

The degenerate case was encountered in most models and only one coefficient,
namely $g^{pp}$, was different from zero. It is then convenient to go back to
the matrix elements $\rho_r(x,x';t)$ by inverting the Fourier 
transform (\ref{5.5e}) so that
the pure decoherence master equation (\ref{5.2e}) becomes
\beq \label{5.19e}
{\partial \over \partial t} \ \rho_r(x, x';t) = - {g^{pp} \over 
\hbar^2} \cdot (x-x')^2 \rho_r(x,x';t) \ .
\eeq

\noi Diagonalization in the basis $\{|x>\}$ is then obvious when 
$g^{pp}$ is a constant since the solution of this equation is
$$\rho_r(x,x',t) = \rho_r (x, x', 0) \exp \left [ - {g^{pp} \over 
\hbar^2} (x-x')^2 t \right ] \ .$$

\noi Similarly, when the only non-zero coefficient is $g^{xx}$, one 
may use the momentum basis $\{|p>\}$ to obtain:
\beq \label{5.20e}
{\partial \over \partial t} \rho_r(p,p';t) = - {g^{xx} \over \hbar^2} 
\cdot (p - p')^2 \rho_r(p,p';t) \ ,
\eeq

\noi implying again diagonalization.\par

A simple condition for the coupling implying diagonalization in the
position basis is given by Eq. (\ref{1.3e}) \cite{34r,15r}. Using
coarse graining, Gell-Mann and Hartle have shown that this condition is
satisfied for mechanical systems when using hydrodynamical observables
as relevant \cite{13r}.

\mysection{The non-degenerate case} \hspace*{\parindent}
Quite different results are obtained in the general case when
there is no degeneracy. One may note first that the differential
operator in the right-hand side of Eq. (\ref{5.16e}) is similar to a
Laplacian, which is given by
\beq \label{6.1e}
\Delta = {1 \over \sqrt{g}} \ {\partial \over \partial x^i} \left ( 
\sqrt{g}g^{ij} {\partial \over \partial^j} \right ) \ ,
\eeq

\noi in the case of a metric $ds^2 = g_{ij} dx^i dx^j$  (with 
$g_{ij}g^{jk} = \delta_i^k$). The factor $g$ is the
determinant of the matrix with elements $g_{ij}$ or the inverse of 
$\det(g^{ij}$). One
could use this remark in principle for a general study of decoherence
but it would need the full power of microlocal analysis. Rather than
entering into such heavy mathematics, it will be convenient to consider
only the case when the coefficients $g^{ij}$ are constants. A further
simplification is obtained by diagonalizing the quadratic form (\ref{5.18e}).
This is done by a change of variables after introducing scale-invariant
parameters: Let $L$ be a unit of ``length'' (i. e. a scale with the
dimensionality of $X$) and $\Pi$ a unit of momentum. The transformation
\beq \label{6.2e}
\Pi X' = \Pi X \cos \theta + L P \sin \theta , LP' = - \Pi X \sin 
\theta + LP co \theta \ ,\eeq

\noi can be viewed either as an ``orthogonal'' change of axes in the $(x, p)$
plane or as a linear canonical transformation. It leaves Weyl's
calculus invariant \cite{25r}, so that if one chooses $\theta$ to 
diagonalize the
metric, one obtains a simpler equation for pure decoherence, namely
(after dropping the prime indices)
\beq \label{6.3e}
{\partial W \over \partial t} = g^{xx} {\partial^2 W \over \partial 
x^2} + g^{pp} {\partial^2 W \over \partial p^2} \ .
\eeq
\vskip 3 truemm

\noi {\large \it General decoherence is not a diagonalization process}\\

The general case of decoherence occurs when the quadratic form (\ref{5.18e})
is non-degenerate. Does then the effect still amount to
diagonalization~? By looking at the degenerate case, one sees
that diagonalization was due to a specific property of the collective
operator $D$: There was a specific orthonormal (``pointer'') basis $\{|j>\}$,
such that
\beq \label{6.4e}
<j|D|j> = 0 \quad \hbox{for each $j$} \ ,
\eeq
\beq \label{6.5e}
{\rm Re} <j|D|k> < 0\ , \quad \hbox{for every pair of indices $j \not= k$} \ .
\eeq

\noi These relations held true for any density matrix $\rho_r$ entering in the
definition of $D$. They must obviously be satisfied if
diagonalization takes place, at least if the basis is independent of
the preparation $\rho_r (0)$ and depends only on the decoherence coefficients.
They do not hold however in general as shown by the following\\

\noi \underbar{No-go theorem} \par

Whatever the state $\psi$, it is impossible for the equation
\beq \label{6.6e}
<\psi |D(\rho_r ) |\psi > = 0
\eeq

\noi to hold for every density matrix $\rho_r$,\\

\noi \underbar{\bf Proof}   According to Eq. (\ref{5.16e}), one can write
\beq \label{6.7e}
D(\rho ) = \Delta \rho \ ,
\eeq

\noi where $\Delta$ is understood as a superoperator acting on a 
collective density $\rho$. One can write
$$<\psi |\Delta \rho|\psi> = Tr_c \left ( |\psi > <\psi |\Delta \rho 
\right ) \ .$$

\noi If this equation is supposed to be valid for any choice of 
$\rho$, one must have (since the superoperator $\Delta$ is hermitian)
$$\Delta |\psi > < \psi | 	  = 0 \ .$$

\noi When written explicitly in the position basis, this equation becomes
$$\left \{ g^{xx} {\partial^2 \over \partial x^2} - \left ( 
g^{pp}/\hbar^2 \right ) \xi^2 \right \} \psi (x + \xi /2) \psi^* (x - 
\xi /2) = 0 \ ,$$

\noi from which one gets
$$\psi (x + \xi /2) \psi^* (x - \xi /2) = a(\xi ) \exp \left ( 
{\sqrt{g^{pp} \over g^{xx}}} (x \xi / \hbar ) \right ) + b(\xi ) \exp 
\left ( - {\sqrt{g^{pp} \over g^{xx}}} (x \xi / \hbar ) \right ) \ 
.$$

\noi This is however impossible (even if the coefficients are
distributions) because it would imply that the wave function of the
state $\psi$ increases exponentially for large values of its argument.\\

\noi \underbar{\bf Note}: The present theorem forbids the existence 
of a universal
diagonalization basis. The possibility of a $\rho$-dependent basis remains
open, although it looks very doubtful.\\

\noi {\large \it Decoherence in the non-degenerate case} \\

Since decoherence cannot be generally a diagonalization process, one
must investigate it anew. Its  consequences are most easily obtained
when $x - x'$ is large. It will be more convenient to use the notation
$(x', x'')$ for the arguments of the reduced density matrix 
$\rho_r(x',x'')$  in the
position representation and to introduce auxiliary variables $x = (x' +
x'')/2$, $\xi = x' - x''$. This means that we are interested on the case where
$\xi$ is large (macroscopic). After performing a Fourier transform to go
back from the variable $p$ to $\xi$, the pure decoherence equation (\ref{5.2e})
becomes
\beq \label{6.8e}
\dot{\rho} = g^{xx} {\partial^2 \rho \over \partial x^2} - \left ( 
g^{pp} \xi^2/\hbar^2 \right ) \rho \ .
\eeq

The time evolution of the function $\rho (x, \xi ) = \rho_r (x + \xi 
/2, x - \xi/2)$ is therefore given by
\beq \label{6.9e}
\rho (x, \xi , t) = \exp \left ( - g^{pp} \xi^2 t / \hbar^2 \right ) 
\times {1 \over \sqrt{4\pi g^{xx}t}} \int dx' \exp \left [ - (x - 
x')^2/4g^{xx}t\right ] \rho (x', \xi , 0 ) \ .
\eeq

The first factor on the right-hand side shows that $\rho_r(x + \xi 
/2, x - \xi /2)$ tends to become
diagonal in the position basis, as in the degenerate case. The heat
kernel in the integral has however a very different effect since it
smoothes off the reduced density along the diagonal, so that
probabilities that were initially distinct become mixed together. If
the process is stopped at some time $t$, its effect is analogous to an
imperfect measurement of the position.\par

The smoothing effect is most clearly seen by considering as initial
state a superposition of two distinct wave functions :
\beq \label{6.10e}
\rho_r (t=0) = |\psi> <\psi|\ , \quad \hbox{with} \ |\psi > = 
|\psi_1> + |\psi_2>
\eeq

\noi the two wave functions $\psi_1(x)$ and $\psi_2 (x)$ being 
clearly separated with clearly
different average values for $X$ or $P$ or both. One is interested
in the interference part of $\rho_r$ originating from $|\psi_1> 
<\psi_2|$ and $|\psi_2> <\psi_1|$ in the initial
state operator, but one must also now consider the probabilistic part
originating from $|\psi_1> <\psi_1|$ and $|\psi_2><\psi_2|$. The 
first factor in the right-hand side of
Eq. (\ref{6.9e}) suppresses very rapidly the interference terms in 
$\rho_r$ when $\psi_1 (x)$ and
$\psi_2 (x)$ have well-separated mean values of the position $X$. It 
also suppresses
them, although less rapidly when the values of $<X>$ coincide while those
of $<P>$ are significantly different \cite{35r}. From the standpoint of
macroscopic interferences, there is therefore nothing new. \par

The smoothing integral in Eq. (\ref{6.10e}) introduces a new
effect. It mixes together the probabilities for different values of
$<X>$. If the state (\ref{6.10e}) represents for instance the state 
of a pointer
after a measurement, two results that would be distinct for
an apparatus with degenerate decoherence can become
indistinguishable if decoherence is non-degenerate. This
conclusion does not depend on the specific form of $\psi_1(x)$ and 
$\psi_2(x)$. It also
holds for coherent states, which are not therefore properly
einselected. \\

\noi {\large \it A symmetric form of decoherence}\\

A convenient expression of non-degenerate decoherence can be obtained
for any number $n$ of collective variables. One denotes altogether by 
$\{\xi^j\}$
the set of the $2n$ position and momentum variables and the pure
decoherence equation becomes
\beq \label{6.11e}
\dot{W} = {\partial \over \partial \xi^i} \left ( g^{ij} {\partial W 
\over \partial \xi^j} \right ) \ .
\eeq

If the decoherence coefficients $g^{ij}$ are constants, one may introduce the
inverse ``covariant'' coefficients $g_{ij}$ satisfying the relations 
$g^{ij} g_{jk} = \Delta_k^i$. They
exist only in the non-degenerate case. The solution of Eq. (\ref{6.11e}) is
then given by
\beq \label{6.12e}
W(\xi t) = (\pi t)^{-n} \int \sqrt{g} d^{2n} \eta \exp \left [ - 
g_{jk} (\xi^j - \eta^j) \left ( \xi^k - \eta^k \right ) /4t \right ] 
W(\eta , 0) \ ,
\eeq

\noi where $g$ is the determinant of the matrix with coefficients 
$g_{ij}$, inverse
of the matrix of the decoherence coefficients $g^{ij}$. It may be useful to
notice that Eq. (\ref{6.12e}) remains approximately valid when the decoherence
coefficients are not constants but slowly varying \cite{25r}.\par

The effect of decoherence is therefore to smear out the Wigner function
in phase space. In this approach, the removal of interference terms is
due to the fact that an interference term, localized in phase space
with a vanishing integral, is rapidly reduced to zero under
smoothing. One may also notice that the results are unchanged under a
linear canonical transformation in phase space, at least when the
coefficients are constants. They are only slightly modified when the
coefficients are slowly varying under a smooth canonical
transformation \cite{25r}.

\mysection{More about degeneracy} \hspace*{\parindent} One may now
consider the order of magnitude of the decoherence coefficients. It
should be stressed first that the condition (\ref{3.1e}) implying
diagonalization was most often imposed {\it a priori} in the construction of
models. If one again considers a unique position observable $X$, only one
decoherence coefficient, $g^{pp}$ , is different from zero when this condition
is satisfied (as shown in Appendix B). Models have revealed a strong
connection between the decoherence coefficient $g^{pp}$  and the friction
coefficient $\gamma^{pp}$, which appears in the classical limit of 
the equation of
motion when the collective Hamiltonian is $H_c  = P^2/2m + V(X)$. 
Classical motion is
then governed by
\beq \label{7.1e}
{dp \over dt} = - \partial V/\partial x - \gamma^{pp} p \ .
\eeq

As shown in Appendix B, the spectral densities of the coefficients $g^{pp}$
and $\gamma^{pp}$  are very similar and, at high enough temperature, the two
coefficients have a simple proportionality relation :
\beq \label{7.2e}  g^{pp} = mT\gamma^{pp} \ .\eeq

\noi What should be considered in that case as a high temperature has been
discussed by Hu, Paz and Zhang \cite{7r}. The fact that $g^{pp}$ enters with a
denominator $\hbar^2$ in the expression (\ref{6.9e}) of decoherence implies
a strong effect of decoherence as soon as $\gamma^{pp}$ is not very
small, i. e. when there is a possibility of dissipation. \\

\noi {\large \it Rough orders of magnitude} \\

Let us consider a model with an environment of oscillators, the
collective system being also an oscillator with frequency $\omega$ [3-8].
One assumes usually a coupling proportional to the collective position
$X$:
$$H_1 = X \cdot \sum_i \left ( \lambda_i a_i^{\dagger} + \lambda_i^* 
a_i \right ) \ ,$$

\noi so that the condition (\ref{1.3e}) is satisfied. More generally, 
one may consider a
coupling with the creation and annihilation operators of the collective
oscillator in place of $X$, i. e.
$$H_1 = (X - iP/m \omega ) \cdot \sum_i \lambda_i a_i + (X + iP/m 
\omega) \cdot \sum_i \lambda_i^* a_i^{\dagger} \ .$$

\noi According to Appendix B, one has then typically:
\beq \label{7.3e}  g^{xx} \approx g^{pp}/m^2 \omega^2 \ .\eeq

\noi According to Eq. (\ref{6.9e}), the decoherence time characterizing the
vanishing of non-diagonal interference terms separated by a distance $\Delta x$
is given as usual by
\beq \label{7.4e}
t_{dec} = {\hbar^2 \over mT \gamma^{pp} \Delta x^2} \ ,\eeq

\noi whereas the characteristic time after which there the probabilities are
mixed up for two different positions on the diagonal separated by the
same distance $\Delta x$ is
\beq \label{7.5e}
t_{mix} \approx {m\omega^2 \Delta x^2 \over \gamma^{pp}T} \ .
\eeq

\noi For reference, it may be recalled that the time necessary for the
spreading of a wave packet on the same distance is
\beq \label{7.6e}
t_{wp} = {m \Delta x^2 \over \hbar} \ .
\eeq

\noi For not too small values of $\Delta x$, and a generic coupling (i. e. no
degeneracy) one has for the various rates: decoherence $\gg$ probability
mix up $\gg$ wave packet spreading. \\

\noi {\large \it On the existence of pointer states} \\

The main conclusion of the previous sections was that the existence of
an exact diagonalization basis is not essential for most physical
consequences of decoherence. On the other hand, it will be now shown
that there is a very large class of physical systems for which such a
basis exists.\par

Coming back to the case when $X$ denotes a class of $n$ collective
coordinates, there is exact diagonalization in the basis $|x>$ if the
derivatives of the symbol $\overline{H}_1$ with respect to the 
canonically conjugate
variables $p$ vanish. One has then according to Eqs. (\ref{5.6e}) and 
(\ref{B.9e}-\ref{B.11e}):
\beq \label{7.7e}
\left [ H_1, X \right ] = 0 \ .\eeq
\beq \label{7.8e}
g^{xp} = g^{xx} = 0 \ .\eeq

Under the same assumptions, according to Eq. 
(\ref{B.17e}-\ref{B.19e}), the friction
coefficients $\gamma^{px}$ and $\gamma^{xx}$ also vanish. \par

When they do not vanish, the classical equations of motion become
\bea
\label{7.9e}
&&dx_i/dt = p_i/m_i - \gamma_{ij}^{px} p_j - \gamma_{ij}^{xx} x_j \ , \\
&&dp_i/dt = F_i - \gamma-{ij}^{pp} p_j - \gamma_{ij}^{px} x_j \ ,
\label{7.10e}
\eea

\noi where $F_i$ denotes a force. \par

These equations look rather unusual and it
is important to understand why they are exceptional (or
unrealistic). In the case of a mechanical system (with no
macroscopic electromagnetic effects) we are familiar
with a unique type of friction coefficient ($\gamma^{pp}$ ) and of decoherence
coefficient ($g^{pp}$). The reason why was clearly shown by Gell-Mann and
Hartle \cite{13r}, who used as position observables the hydrodynamical
variables resulting from a coarse graining. The
corresponding variables can be identified with the center-of-mass 
positions $x_i$ of
small pieces of matter, small enough from a macroscopic standpoint
although containing a large number of atoms. The key feature is then
the non-relativistic form of the Hamiltonian for the
particles of matter,
$$H = \sum \left ( p_{\alpha} - A(x_{\alpha})\right )^2 /2m_{\alpha} 
+ \sum V\left ( x_{\alpha} - x_{\beta} \right ) \ ,$$

\noi where the summations are performed over the particles (indicated by
Greek indices). If there is no macroscopic magnetic field (so that one
can neglect the magnetic potential $A$), one of the Heisenberg equations
of motion, yields the following simple relation between the
classical velocity and momentum (denoted by Latin indices)
$$\dot{x}_i = p_i/m_i \ .$$

\noi Comparing this with Eq. (\ref{7.9e}), one sees that $\gamma^{px} 
= \gamma^{xx} = 0$ and, from
Eq. (\ref{B.16e}), one may expect that $g^{xp} = g^{xx} = 0$. Eqs. 
(\ref{7.9e}-\ref{7.10e}) strongly suggest that
this property follows from the Galilean invariance of non-relativistic
mechanics under a change of reference system.\par

Strangely enough, no realistic example of the non-degenerate case has
yet been proposed, except tacitly in unprecise measurements. Examples 
might be expected however
in electromagnetic systems (where the magnetic and electric fields
replace position and conjugate momentum) but the prospect of producing
quantum superpositions of fields and see their decoherence seems
rather remote. One must probably attribute the rarity of examples to
the fact that decoherence has been mostly studied in the framework of
measurement theory. There is almost always (or always) in that case
some mechanical part of some apparatus that is entangled with the
measurement result and the rest of the system, and it enforces its
own einselection on them.

\mysection{Classical behavior}
\noi {\large \it The derivation of classical behavior} \\

      Decoherence is in most cases
immediately followed by a classical behavior of the collective
subsystem \cite{4r,10r,12r,23r}. Although this property will not be analyzed
in detail in the present work, a few points involving again the
problem of einselection are worth mentioning.\par

Decoherence is described in the non-degenerate case by Eq. (\ref{6.12e})
involving a smearing effect on the Wigner function $W(x, p)$. An important
consequence is to make this function non-negative so that its
interpretation as a density probability in phase space becomes
significant \cite{36r}. As far as orders of magnitude are concerned, one may
consider that a derivative operator $\partial / \partial x$ acting on 
$W$ is of the order of
$(g^{xx}t)^{-1/2}$ for $t$ large enough (i. e. when decoherence is 
effective) whereas $\partial /\partial p$
is of the order of $(g^{pp}t)^{-1/2}$. One may then
consider more carefully the first term in the master equation (\ref{4.6e})
giving the following contribution to the master equation
\beq \label{8.1e}
\dot{\rho}_r = -{i \over \hbar} \left [ H_c, \rho_r \right ] \ .\eeq

One can write down this equation in terms of the Wigner function and
the Hamilton function $h(x, p)$, which are respectively the symbols of $\rho_r$
and of $H_c$. It reads to second order in $\hbar$ :
\beq \label{8.2e}
{\partial W \over \partial t} = -{\partial h \over \partial p} 
{\partial W \over \partial x} + {\partial h \over \partial x} 
{\partial W \over \partial p} + {\hbar^2 \over 24} \left \{ 
{\partial^3h \over \partial p^3} {\partial^3W \over \partial x^3} - 3 
{\partial^3h \over \partial x \partial p^2} {\partial^3W \over 
\partial p \partial x^2} + 3 {\partial^3h \over \partial p \partial 
x^2} {\partial^3W \over \partial x \partial p^2} - {\partial^3h \over 
\partial x^3} {\partial^3W \over \partial p^3} \right \} \ .
\eeq

\noi One recognizes in the first term in the right-hand side a Poisson
bracket of the Hamilton function and the probability density, in
agreement with classical physics. This term generates a
classical evolution of the Wigner function as if its arguments $(x, p)$
were moving according to the classical Hamilton equations.

The order of magnitude of the $W$-derivatives resulting from Eq. (\ref{6.12e})
imply that higher order corrections in $\hbar$ are negligible so that after
some decoherence the evolution becomes classical. It is
somewhat paradoxical that the analysis is more involved in the
degenerate case \cite{10r}. The difficulty arises from a linear
superposition of two initial wave functions with the same
$x$-location but different average values of $p$. The destruction of
interferences must then wait till the motion due to the difference in the
values of $<P>$ separates the wave functions in $x$-space. A conspiracy of
decoherence with the collective dynamics is therefore necessary  for
producing finally a classical behavior. \\

\noi {\large \it Classicality and the choice of a collective subsystem} \\

      The previous conclusion of a
classical behavior assumed tacitly that the derivatives of the Hamilton
functions are not large, but one might then get involved in a
circular argument. The collective observables are chosen ordinarily on
empirical grounds, from a direct knowledge of the system. One says: ``I
look at the system and I clearly see how it can be described by some
coordinates, which I replace by quantum
observables''. Then one concludes after much work: ``See! The description
of the system with these variables becomes finally classical''. This is
certainly not a proof of classicality resting on the basic principles
of quantum mechanics, but only a check of consistency: Classical
behavior can be proved when the convenient variables for describing it
have been selected by means of one's classical intuition.\par

The question ``How does one select a collective subsystem?'' is therefore
prior to the question of einselection. One can then look at Eq. 
(\ref{8.2e}) from a different standpoint: It
should give a criterion for choosing the collective observables and 
not provide a
proof that they describe a classical motion. This criterion implies 
that the derivatives $\partial/\partial x$ and $\partial / \partial 
p$ of
the classical Hamilton function in the second and higher terms
of Eq. (\ref{8.2e}) are not controlled by factors involving some 
power of $\hbar^{-1}$.\par

This gross criterion can be presumably much refined in view of
Fefferman's formulation of quantum mechanics through
pseudo-differential calculus (microlocal analysis) \cite{37r}. He
investigated the eigenstates of the complete hamiltonian $H$ of an
arbitrary quantum system by analyzing its symbol $\overline{H}$ in 
the phase space
of the constituent particles and, by cutting this space into
``distorted boxes'', he was able to diagonalize $H$
approximately. This is a deep result of abstract mathematics but there has
been no direct application of it in physics.
Nevertheless, it means that there exists one (or
several) privileged ways of cutting phase space, into well-defined 
boxes, according
to the possible states of the system. One could then envision that
a pair of variables $(x, p)$ is collective if it defines locally a
2-dimensional plane along which $\overline{H}$ varies very slowly. 
Such a property
is strongly suggested by Feffermann's construction and it agrees with the
small derivatives we just found characterizing classical
behavior after decoherence. One may also presume the existence of a
whole hierarchy of collective 2-directions, which would be ordered
according to the magnitude of the derivatives.\par

I will not try to elaborate further on this idea, which was
proposed some years ago although not much progress has been made since
\cite{26r}. It represents however an alternative to Zurek's
predictability sieves (with which it may be related). In any case,
it stresses again that the most important problem in a real
understanding of decoherence is an explicit construction of the collective
observables (with a corresponding explicit definition of the
environment). This is closely related with a search for a
real theory of the Heisenberg frontier, as also noticed by Zurek \cite{10r}. \\

\noi {\large \it Einselection} \\

       Zurek's concept of predictability sieves was applied successfully to
the case of an underdamped collective oscillator interacting with an
environment of oscillators [17,40-43]. It was found that Gaussian pure
states with various average values $(x, p)$ of position and momentum are
selected in that case as the best carriers of information, suggesting
more generally that some sort of coherent states would be einselected
by decoherence just before classical motion. There is
something puzzling however in the fact that the width of these Gaussian states
is controlled by the parameters $(m, \omega)$ of the collective oscillator
(it has the same width as the ground state wave function of the
oscillator). When looking at Eq. (\ref{6.12e}), one finds on the contrary that
decoherence in the non-degenerate case is insensitive to the
characteristics of the collective Hamiltonian and it is completely determined
by the coupling Hamiltonian through the decoherence coefficients. \par

It may be recalled in this connection that a convenient family of
einselected states was proposed earlier, although this name was
not used \cite{38r,39r}. These states are closely related to 
H\"ormander's notion of
microlocal projection operators. The symbol $\overline{P}(x,p)$ of 
such an operator is
zero outside a regular cell $C$ in the $(x, p)$ phase space (i. e.
a cell whose volume and boundary shape have large characteristic
dimensions in terms of the Planck constant). $\overline{P}(x,p)$ is equal to
1 in $C$, except near the boundary where it goes smoothly to zero. 
The corresponding operator $P$ is practically a
projection \cite{25r}.\par

Consider then the integrand of Eq. (\ref{6.12e}) for definite values 
of $t$ , $\xi =
(x, p)$ and $\eta = (x', p')$
\beq \label{8.3e}
\overline{Z} (\xi ) = (\pi t)^{-n} \sqrt{g} \exp \left [ - g_{jk} 
\left ( \xi^j - \eta^j \right ) \left ( \xi^k - \eta^k \right ) /4t 
\right ] \ .
\eeq

\noi It can be considered as the symbol of a density operator $Z$ 
originating at
a time $t$ large enough through non-degenerate decoherence from an initial
state localized in the neighborhood of $\eta$. Using Eq. 
(\ref{5.6e}), one finds
that $PZ = Z$ if the cell $C$ contains a manifold with equation
$$g_{jk} \left ( \xi^j - \eta^j \right ) \left ( \xi^k - \eta^k 
\right )/4t = a$$

\noi for a large enough value of $a$. (This property remains valid 
when the decoherence coefficients
$g^{ij}$ are not constant but smoothly varying).\par

This means that the normalized mixed states with density matrix $P/TrP$
satisfy the criteria for einselected states. This includes their
sifting through predictability sieves \cite{10r}, since $Tr\Xi^2(t)
\approx 1$ for a reduced density matrix $\Xi$ such that $\Xi (0) =
P/TrP$. The sketch of the proof consists in separating diagonalization
and mixing according to Eq. (6.9) through a canonical transformation
maximizing the rate of diagonalization. The sifting property follows
when $\underbar{t}$ is such that diagonalization has already taken
place in the cell $C$ whereas mixing has not spilled outside $C$. This
is valid for non-degenerate and degenerate decoherence.\par

One can then identify einselected states in general with the classically
meaningful states, which are defined either as classical properties
through the projections $P$ \cite{39r,44r} or as quantum states by 
the density operators
$P/TrP$. The predictability sieve criterion is universally valid.
Its stability under a change of definition for the collective subsystem
and the environment (i. e. under a shift of Heisenberg's frontier)
cannot be proved however along the same lines as long as no objective 
definition of the
collective observables has been found.

\mysection{Conclusions} \hspace*{\parindent}
As suggested by its name, decoherence is a loss of correlation between
local phases of a system involving a large number of constituents. It
may take in principle many different aspects because ``in principle'' the
set of states of a quantum system is enormous, even much more than its
Hilbert space. Empirical physics is however interested in the
systems really occurring in nature or built in the laboratory,
which can be measured or observed.\par

A wide gap between theory and practice is our unability
to characterize mathematically these ``real'' systems \cite{12r}. 
There is a wide agreement that they always
involve some ``collective'' degrees of freedom but the problem of 
their definition from first principles is not yet solved. The study 
of decoherence
will probably remain semi-empirical as long as the program suggested 
in Section 8, or an
equivalent one, is not completed.\par

The practical results of the present study were concerned with the three
main aspects of decoherence: suppression of macroscopic
interferences, einselection  and later classical behavior.\par

The suppression of macroscopic interferences is a general feature. 
The interference terms disappear for two collective wave functions
with a large enough difference in the average values of position
or momentum (or both).\par

Einselection is the election of definite states representing exclusive
events with well-defined probabilities.
It is essential in measurement theory and its
properties were the main purpose of this paper. Two different cases 
had to be distinguished, which were respectively called
degenerate and non-degenerate. \par

There is something puzzling in this dichotomy if one does
not distinguish also between what is most general (or frequent) either from
the standpoint of a mathematical theory or of
empirical physics. A very large class of
physical systems leads to the degenerate form of einselection, which is
practically a diagonalization of the reduced density matrix in the
basis originating from the collective position
coordinates. These systems are truthfully
described by hydrodynamical variables after coarse graining \cite{12r,13r}.
Although this condition is still restrictive, it turns out in practice
that the mechanical parts of a physical system, which are described by
these variables are entangled most often with other degrees of
freedom so that degeneracy (with diagonalization) is extended to
them. A simpler way of saying this is that most
observations and measurements involve or could involve a reading of the
position of some mechanical ``pointer'', imposing diagonalization as
the outcome of decoherence.\par

A sufficient condition for degeneracy is given by the well-known Eq. 
(\ref{1.3e}), which covers the hydrodynamical case. It is very
restrictive however from a mathematical standpoint and, in the absence of a
criterion defining a realistic system, one had also to investigate the
general case of non-degenerate decoherence. The results did
not quite agree with the conjecture of predictability
sieves \cite{10r}. One found a tendency of decoherence to combine an
approximate diagonalization with a partial lumping of
probabilities rather than a clear mutual exclusion of events, which 
is typical of unprecise measurements.

The situation was clearer when one looked at the classical behavior
after decoherence. There is a simple way to
reconcile the present results with the Zurek's predictability 
criteria \cite{10r}. It consists in identifying the einselected
states with the mixed states representing classical
properties, which I proposed earlier \cite{44r}. The fact that these 
states are best defined by the mathematics of
microlocal analysis \cite{25r}, as well as Fefferman's promising approach to
the definition of collective observables \cite{37r}, indicate in my
opinion that this framework is the right one.

\newpage
\appendix
\mysection{- Appendix A. Decoherence as an irreversible\break\noindent process}

\noi {\large \it Auxiliary densities}\\

      The simplest way for computing the quantities $s_i$ for decoherence
consists in using the observables $(|k><k'| + |k'> <k|) \otimes I_e$
and $(1/i)(|k><k'| - |k'> <k|) \otimes I_e$ (for $k \not= k'$, $k >
k'$) together with the diagonal terms $|k><k|\otimes I_e$. They will be
denoted altogether by $A^j$ and they satisfy the relations
\beq \label{A.1e}
Tr_c A^j A^{j'} = \delta^{jj'} \quad (j,j' \not= 1, e) \ .
\eeq

\noi Rather than, the exponential form (\ref{2.4e}) for the test 
density operator,
it is convenient to write it as
\beq \label{A.2e}
\rho_0 = \left ( \sum_{j\not= 1,e} a^j \ A^j \right ) \otimes \rho_e \ .
\eeq

\noi From Eqs. (\ref{A.1e}), one sees that the coefficients $a^j$ in 
Eq. (\ref{A.2e}) are the
average values of the observables $A^j$. Therefore
\beq \label{A.3e}
\partial \rho_0 / \partial a^j = A^j \otimes \rho_e \quad (j \not= 1, e) \ .
\eeq

\noi After writing $\rho_0 = \exp (- \alpha - \beta H_e)$, one obtains
\beq
\label{A.4e}
\partial \beta / \partial E = - \Delta^2 \ , \  \partial \alpha / 
\partial E = E/\Delta^2 \ , \ \partial \rho_e / \partial \alpha = - 
\rho_e \ , \ \partial \rho_e / \partial \beta = - H_e \rho_e
\eeq

\noi (where $\Delta$ is defined by Eq. (\ref{3.8e})). In view of the
definition (\ref{2.5e}) for the auxiliary densities (or equivalently the
definition (\ref{2.8e}) of the projection $P$), their expression 
(\ref{3.5e}-\ref{3.9e}) in
Section~3 follows immediately from Eqs. (\ref{A.3e}-\ref{A.4e}). \\

\noi {\large \it The evolution equations} \\

In the first evolution equation (\ref{2.13e}), one can compute 
$PL\rho_1$ by applying
the expression (\ref{3.9e}) giving $P$ to $L\rho_1 = - (i/\hbar ) [H, 
\rho_1]$. Two traces $tr [H, \rho_1]$ and $Tr(H_e[H,\rho_1])$ enter 
in the
result. Using cyclic invariance of traces together with Eq. (\ref{3.10e})
specifying $H_1$ and
\beq
\label{A.5e}
\rho_1 = \rho_r \otimes \rho_e \ ,
\eeq

\noi one finds that
\begin{eqnarray*}
&&tr \left [ H, \rho_1 \right ] = \left [ H_c , \rho_r \right ] \ , \\
&&Tr \left ( H_e\left [ H, \rho_1 \right ] \right ) = Tr \left ( 
\left [ H_e, H \right ] , \rho_1 \right ) = Tr \left ( \left [ H_e, 
H_1 \right ] , \rho_1 \right ) \ .
\end{eqnarray*}

\noi Therefore
\beq
\label{A.6e}
PL\rho_1 = (-i/\hbar) \left \{ \left [ H_c , \rho_r \right ] \otimes 
\rho_e + I_c \otimes \rho_e {H_e - E \over \Delta^2} Tr \left ( H_e 
\left [ H_1 , \rho_1 \right ] \right ) \right \} \ .\eeq

In order to compute $\dot{P}\rho_1$, one remarks that although the 
quantities $\rho_e$, $E$
and $\Delta$ in Eq. (\ref{3.9e}) are time-dependent, the quantity 
$Tr(H_e\rho_1) - ETr\rho_1$ vanishes so that
one has
\beq \label{A.7e}
\dot{P} \rho_1 = \rho_r \otimes \dot{\rho}_e - \dot{E}I_c \otimes 
\rho_e {H_e - E \over \Delta^2} \ .
\eeq

\noi An identical result is obtained for $\dot{P}\rho$, so that
\beq \label{A.8e}
\dot{P}\rho_2 = 0 \ .\eeq

One thus get the first evolution equation
\beq \label{A.9e}
\dot{\rho}_1 = PL\rho_1 + \dot{P}\rho_1 + PL \rho_2 \ ,
\eeq

\noi with the expressions (\ref{A.6e}) and (\ref{A.7e}) for the first 
two terms whereas $PL\rho_2$ is
given by Eq. (\ref{3.9e}).

It is convenient to split this equation into one for $\dot{\rho}_r$ 
and another for
$\dot{\rho}_e$ (or equivalently for $\dot{E}$). This is obtained by taking respectively the
trace of Eq. (\ref{A.9e}) with respect to the enviro 
nment and the collective
Hilbert space. The environment trace of the second term in the
right-hand side of Eq. (\ref{A.6e}) vanishes, as well as 
$tr\dot{P}\rho_1$ (because $tr\dot{\rho}_e = 0$).
According to Eq. (\ref{3.9e}), the environment trace of $PL\rho_2$ reduces to
$$- {i \over \hbar} tr \left [ H, \rho_2 \right ] = - {i \over \hbar} 
tr \left [ H_1, \rho_2 \right ] \ ,$$

\noi where the second equality results from the vanishing of 
$tr[H_e,\rho_2]$ (as the trace
of a commutator) and of $tr[H_c, \rho_2]$ (because of $tr\rho_2 = 
0$). One obtains thus the
basic equation
\beq \label{A.10e}
\dot{\rho}_r - {i \over \hbar} \left ( \left [ H_c, \rho_r\right ] + 
tr \left [ H_1, \rho_2 \right ] \right ) \ .
\eeq

\noi The trace of Eq. (\ref{A.9e}) on the collective Hilbert space reduces to a
(potentially infinite) term $(Tr_cI_c)\rho_e (H_e - E)$, multiplied 
by a number, which must then
vanish so that
\beq \label{A.11e}
\dot{E} = - {i \over \hbar} Tr \left ( H_e\left [ H_1, \rho_1 + 
\rho_2 \right ] \right ) = - {i \over \hbar} Tr \left ( H_e \left [ 
H_1, \rho_2 \right ] \right ) \ ,
\eeq

\noi the last equality resulting from $Tr(H_e[H_1,\rho_1 ] = Tr (H_1 
[H_e , \rho_1 ]$, whereas
\beq
\label{A.12e}
\left [ H_e , \rho_1 \right ] = 0 \ ,
\eeq

\noi since $\rho_e$ is a function of $H_e$. \\

\noi {\large \it The second evolution equation} \\

One can now write down the evolution equation (\ref{2.14e}) for 
$\rho_2$, which is formally.
\beq
\label{A.13e}
\dot{\rho}_2 = QL\rho_2 + QL\rho_1 - \dot{P}\rho_1
\eeq

\noi (after taking Eq. (\ref{A.8e}) into account). This will be done
according to Section 4 by considering $H_1$ as a perturbation. One needs
only to compute $\rho_2$ at first order in $H_1$ and some terms in 
Eq. (\ref{A.13e}) can be
therefore immediately neglected. For instance $\dot{P}\rho_1$, as 
given by Eq. (\ref{A.7e})
is negligible because $\dot{E}$ is of second order (according to Eq. 
(\ref{A.11e}) and
furthermore,
\beq
\label{A.14e}
\dot{\rho}_e = - \dot{\beta} \left ( H_e - E\right ) \rho_e = \dot{E} 
\Delta^{-2} \left ( H_e - E \right ) \rho_e
\eeq

\noi is also of second order (the second equality resulting from 
Eq.(\ref{A.4e}). One can also neglect the commutator $[H_1 , \rho_2]$ 
in $L\rho_2$ as being of second order.\\

Let us now consider the quantity ${\cal Q}L\rho_1 = L \rho_1 - PL 
\rho_1$. One has
\beq
\label{A.15e}
L \rho_1 = - (i/\hbar ) \left [ H , \rho_1 \right ] = - (i/\hbar ) 
\left [ H_c + H_1 , \rho_1 \right ] \ ,
\eeq

\noi where the second equality results from Eq. (\ref{A.12e}). Then
\beq
\label{A.16e}
PL\rho_1 = tr \left ( L \rho_1 \right ) \otimes \rho_e + \Delta^{-2} 
\left ( I_c \otimes H_e - E \right ) \rho_1 \left \{ tr \left ( I_c 
\otimes H_e L \rho_1 \right ) - ETr \left ( L \rho_1 \right ) \right 
\} \ .
\eeq

\noi The last term $Tr(L\rho_1)$ vanishes (as a trace of a 
commutator). The preceding term also vanishes since
$$Tr \left ( I_c \otimes H_e L \rho_1 \right ) = - (i/\hbar ) Tr 
\left ( I_c \otimes H_e \left [ H, \rho_1 \right ] \right ) = (i/ 
\hbar)Tr \left ( H \left ( I_c \otimes H_e, \rho_1 \right ] \right ) 
	$$

\noi and the last commutator vanishes. Therefore
$$PL \rho_1 = (-i/ \hbar ) tr \left [ H, \rho_1 \right ] \otimes 
\rho_e = (-i/ \hbar ) \left [ JH_c, \rho_r \right ] \otimes \rho_e  = 
(-i/\hbar) \left [ H_c \otimes I_e, \rho_1 \right ] \ ,$$

\noi where the second equality results from Eqs. (\ref{3.10e}) and
(\ref{A.12e}). The first term in $L\rho_1$ as given by Eq. (\ref{A.15e}) is
therefore cancelled and one is left with the simple equation
\beq
\label{A.17e}
\dot{\rho}_2 = - (i/\hbar ) \left [ H_0, \rho_2 \right ] - (i/\hbar ) 
\left [ H_1 , \rho_1 \right ]
\eeq

where $H_0$ is the uncoupled hamiltonian:
\beq
\label{A.18e}
H_0 = H_c \otimes I_e + I_c \otimes H_e \ .
\eeq

\newpage

\mysection{- Appendix B. Decoherence and dissipation}
\noi {\large \it Derivation of Eq. (\ref{5.8e})}\\

According to Eq. (\ref{5.2e}), one must evaluate at lowest order in 
$\hbar$ the symbol $\overline{D}$ of the collective operator
\beq
\label{B.1e}
D = - (1/\hbar )^2 tr \left [ H_1, \left [ H_1^T, U(t-t') \rho_r (t') 
\otimes \rho_e (t') U^{-1} (t - t') \right ] \right . \ .
\eeq

\noi Since $U(t) = U_c (t) \otimes U_e(t)$ (with $U_c(t) = \exp 
(-iH_c t/\hbar )$ and $U_e(t) = \exp (-iH_et/\hbar))$, one can 
slightly simplify the density operator by writing
\beq
\label{B.2e}
U(t-t') \rho_r(t') \otimes \rho_e(t') U^{-1}(t-t') = U_c(t-t') \rho_r 
(t') U_c^{-1} (t-t') \otimes \rho_e(t') \ ,
\eeq

\noi in view of the equality (resulting from Eq. (\ref{A.14e}) and 
valid up to order $H_1^2$)
$$U_e(t-t') \rho_e(t') U_e^{-1}(t-t') \approx \rho_e(t') \ .$$

Letting $A$ and $B$ be two arbitrary operators, $\overline{A}(x,p)$ and
$\overline{B}(x,p)$ their operator-valued symbols, Eq. (\ref{5.5e}) gives the
symbol of the commutator $[A, B]$, with the notation of Section~5:
\beq \label{B.3e}
(-i\hbar /2) \left ( \overline{A}_p \overline{B}_x - \overline{A}_x 
\overline{B}_p - \overline{B}_p \overline{A}_x + \overline{B}_x 
\overline{A}_p \right ) + O(\hbar^3) \ ,
\eeq

\noi Eq. (\ref{B.3e}) can be used twice for obtaining the symbol of the double
commutator in Eq. (\ref{B.1e}). The symbol of the operator $U_c(t-t') 
\rho_r (t') U_c^{-1}(t-t')$, which will be
denoted by $W^T$ is an ordinary function and it commutes with the
operator-valued symbols $\overline{H}_1, \overline{H}_1^T$ and their 
derivatives. After a
straightforward calculation, one gets:
\beq \label{B.4e}
\overline{D} = \int_{-\infty}^t dt' \left ( {\partial \over \partial 
x} \left ( C^{xx} W_x^T + C^{xp} W_p^T \right ) + {\partial \over 
\partial  p} \left ( C^{px} W_x^T + C^{pp} W_p^T \right ) \right ) \ 
.\eeq

\noi The coefficients are given by
\bea
\label{B.5e}
&&C^{xx} = {1 \over 2} tr \left \{ \left ( \overline{H}_p 
\overline{H}_p^T + \overline{H}_p^T \overline{H}_p \right ) \rho_e 
\right \} \ , \\
&&C^{xp} = - {1 \over 2} tr \left \{ \left ( \overline{H}_p 
\overline{H}_x^T + \overline{H}_x^T \overline{H}_p \right ) \rho_e 
\right \} \ , \\
\label{B.6e}
&&C^{px} = - {1 \over 2} tr \left \{ \left ( \overline{H}_x 
\overline{H}_p^T + \overline{H}_p^T \overline{H}_x \right ) \rho_e 
\right \} \ , \\
\label{B.7e}
&&C^{pp} = {1 \over 2} tr \left \{ \left ( \overline{H}_x 
\overline{H}_x^T + \overline{H}_x^T \overline{H}_x \right ) \rho_e 
\right \} \ ,
\label{B.8e}
\eea
\vskip 3 truemm

\noi {\large \it Explicit expressions} \\

The decoherence coefficients after neglecting retardation are obtained
by integrating the coefficients (\ref{B.5e}-\ref{B.8e}) on the time $t'$. It is
convenient to introduce the matrix elements of $\overline{H}_1$~:
$$<n|\overline{H}_{1i}^T|n'> = \overline{H}_{1inn'} \exp \left ( - i 
\omega_{nn'} \tau \right )$$

\noi where the index $i$ denotes either $x$ or $p$, $\tau = t - t'$ 
and $\overline{H}_{1inn'} = <n|\overline{H}_{1i}|n'>$. One has
then
$${1 \over 2} tr \left \{ \left ( = \overline{H}_i \overline{H}_j^T + 
\overline{H}_j^T \overline{H}_i \right ) \rho_e \right \} = 
\sum_{nn'} \overline{H}_{1inn'} \overline{H}_{1jn'n} \exp \left ( i 
\omega_{nn'} \tau \right ) p_{nn'} \cosh \left ( \beta \hbar 
\omega_{nn'} /2 \right ) \ .$$

\noi The decoherence coefficients
$$g^{ij} = \int_0^{\infty} d\tau C^{ij} (\tau )$$

\noi are then given by
\bea
\label{B.9e}
&&g^{xx} = \int_0^{\infty} d\tau \sum_{nn'} \overline{H}_{1pnn'} 
\overline{H}_{1pn'n} \exp \left ( i \omega_{nn'} \tau \right ) 
p_{nn'} \cosh \left ( \beta \hbar \omega_{nn'} /2 \right ) \ , \\
&&g^{px} = g^{xp} = - \int_0^{\infty} d\tau \sum_{nn'} 
\overline{H}_{1pnn'} \overline{H}_{1xn'n} \exp \left ( i \omega_{nn'} 
\tau \right ) p_{nn'} \cosh \left ( \beta \hbar \omega_{nn'} /2 
\right ) \ , \\
\label{B.10e}
&&g^{pp} = \int_0^{\infty} d\tau \sum_{nn'} \overline{H}_{1xnn'} 
\overline{H}_{1xn'n} \exp \left ( i \omega_{nn'} \tau \right ) 
p_{nn'} \cosh \left ( \beta \hbar \omega_{nn'} /2 \right ) \ .
\label{B.11e}
\eea

\noi The quadratic form in two real variables $(\alpha , \alpha ')$ 
with these coefficients is given by
\beq
\label{B.12e}\ .
g^{xx} \alpha^2 + 2g^{xp} \alpha \alpha ' + g^{pp} \alpha '^2 = 
\int_0^{\infty} d\tau \sum_{nn'} \left | \overline{H}_{1xnn'} \alpha 
- \overline{H}_{1pn'n} \alpha ' \right |^2\cos \left ( \omega_{nn'} 
\tau \right ) p_{nn'} \cosh \left ( \beta \hbar \omega_{nn'} /2 
\right )
\eeq

\noi and it is clearly non-negative. \\

\noi {\large \it Dissipation coefficients}\\

Let now $A_k$ ($k = 1$ or 2) denote either $X$ or $P$. The time 
derivatives of their average values are given by
\beq
\label{B.13e}
{d<A_k> \over dt} = {i \over \hbar} Tr \left ( \left [ H, A_k \right 
] \rho \right ) = {i \over \hbar} Tr_c \left ( \left [ H_c, A_k\right 
] \rho_r \right ) + {i \over \hbar} Tr \left ( \left [ H_1 , A_k 
\right ] \rho_2 \right ) \ .
\eeq

\noi The first term is due to collective dynamics and the second one 
represents dissipation effects. Using Eq. (\ref{4.5e}) for $\rho_2$, 
this dissipative term reads
\beq
\label{B.14e}
- {i \over \hbar^2} \int_0^{\infty} d\tau Tr \left \{ \left [ H_1, 
A_k \right ] \left [ H_1^T, U(\tau ) \rho_r \otimes \rho_e U^{-1} 
(\tau ) \right ] \right \} \ .
\eeq

\noi Eq. (\ref{5.7e}) can be used to replace the collective trace by 
an integration over phase space of the operator symbol. One can use
$$\overline{[H_1, X]}  = - i \hbar \overline{H}_{1p}, \overline{[H_1, 
P]}  = i\hbar \overline{H}_{1x} \ .$$

\noi When computing the symbol of $[H_1^T, U(\tau ) \rho_r \otimes 
\rho_e U^{-1}(\tau )]$, one will retain only the term
originating from the Poisson bracket between $H_1^T$ and either 
$U(\tau)$ or $U^{-1}(\tau)$,
because it can be seen that all the other contributions do not
contribute to the phase space integral or are of higher order in $\hbar$.
Finally, denoting by $ \overline{H}_c(x,p)$ the symbol of $H_c$ 
(which is the collective
Hamilton function), one gets
$$\overline{[H_1^T, U(\tau) \rho_r \otimes \rho_e U^{-1}(\tau )]} = 
{i \hbar \over 2} \left \{ \left ( \overline{H}_{1p}^T \rho_e - 
\rho_e \overline{H}_{1p}^T \right ) V'(x) - \left ( 
\overline{H}_{1x}^T \rho_e - \rho_e \overline{H}_{1x}^T \right ) {p 
\over m} \right \} \tau W .$$

\noi In the semi-classical case, when the Wigner function is slowly varying,
Eq. (\ref{B.13e}) generates the classical equations of motion, which read
(after writing $(p, x)$ in place of $<P>$, $<X>$ and neglecting retardation):
\bea
\label{B.15e}
&&{dp \over dt} = - \overline{H}_{cx} - \gamma^{pp} \overline{H}_{cp} 
- \gamma^{px} \overline{H}_{cx} \ ,\\
&&{dx \over dt} = - \overline{H}_{cp} - \gamma^{xp} \overline{H}_{cp} 
- \gamma^{xx} \overline{H}_{cx} \ .
\label{B.16e}
\eea

The coefficients are explicitly given (after performing a partial 
integration over the time $\tau$) by
\bea
\label{B.17e}
&&\gamma^{pp} = \int_0^{\infty} d\tau \overline{H}_{1xnn'} 
\overline{H}_{1xn'n} \exp \left ( i \omega_{nn'} \tau \right ) 
p_{nn'} \left ( \sinh \left ( \beta \hbar \omega_{nn'}/2 \right ) / 
\hbar \omega_{nn'} \right ) \ , \\
&&\gamma^{px} = - \int_0^{\infty} d\tau \overline{H}_{1xnn'} 
\overline{H}_{1pn'n} \exp \left ( i \omega_{nn'} \tau \right ) 
p_{nn'} \left ( \sinh \left ( \beta \hbar \omega_{nn'}/2 \right ) / 
\hbar \omega_{nn'} \right ) = \gamma^{xp} , \\
\label{B.18e}
&&\gamma^{xx} = \int_0^{\infty} d\tau \overline{H}_{1pnn'} 
\overline{H}_{1pn'n} \exp \left ( i \omega_{nn'} \tau \right ) 
p_{nn'} \left ( \sinh \left ( \beta \hbar \omega_{nn'}/2 \right ) / 
\hbar \omega_{nn'} \right ) \ .
\label{B.19e}
\eea

\noi Comparing these results with Eq. (\ref{B.9e}-\ref{B.11e}), a
strong formal similarity appears between the dissipation and the
decoherence coefficients. They are even directly proportional at high
enough temperature (when $T \gg \hbar \Omega$ so that $\cosh (\beta 
\hbar \omega_{nn'}/2) \approx 1$ and $\sinh (\beta \hbar 
\omega_{nn'}/2) \approx \beta \hbar \omega_{nn'}/2)$,
namely:
\beq \label{B.20e}
\gamma^{il} \approx g^{ij}/T \ .
\eeq

\noi One may also notice that the dissipation of collective energy $dE_c/dt$ is
always negative (or zero), since the quadratic form with coefficients 
$\gamma^{ij}$
is non-negative for the same reason as in Eq. (\ref{B.12e}). Finally, the
equality $\gamma^{px} = \gamma^{xp}$ is a special case of the Onsager 
symmetry relations (24).

\newpage
\mysection{- Appendix C. Previous models}
\hspace*{\parindent}
Much of our knowledge on decoherence was first obtained from a study of
simple models. Since the present theory claims a much wider range, it
should at least recover these older results. This will be the topic of
the present Appendix.\\

\noi {\large \it Oscillator bath} \\

A model of environment consisting of a bath of harmonic oscillators has
been much investigated. For an atom interacting with radiation in a
cavity, the model is exact and its results have been successfully
compared with experiment \cite{1r}. The atom is represented in that case by a
two-state system. When considering a collective system with position $X$,
the coupling Hamiltonian is typically written as
\beq \label{C.1e}
H_1 = X \cdot \left ( \sum_i \left ( \lambda_i a_i + \lambda_i^* 
a_i^{\dagger} \right ) \right ) \ ,
\eeq

\noi where the sum is over all the oscillators, $\lambda_i$ being a 
coupling constant,
  $a_i$ and $a_i^{\dagger}$ their annihilation and creation operators. \\

It is easy to use this coupling in the formulas of Appendix B and to
recover the previously known results. The calculations are essentially
trivial and need not be given here explicitly. Perhaps more interesting
is the question of the range of this model. In addition to their
thorough study of it, Caldeira and Leggett suggested that it should be
considered as very general \cite{5r}. They start from the fact that the
number of energy eigenstates $|n>$ of the environment is extremely large.
They introduce formally an oscillator for each such state (i.e. the
index $i$ will stand for $n$ in Eq. (\ref{C.1e})). They notice that a 
state $|n>$ is
either occupied or not, these two possibilities being represented by
the ground state and the first excited state of the associated
oscillator. Then they argue that higher excited states of the
oscillators will play no role because of the smallness of their
probability of excitation. They conclude that an oscillator bath can
represent almost any environment. \\

The argument is however erroneous, because the coupling resulting from
their proposal would not have the very simple form of Eq. (\ref{C.1e}) in
general, because transitions $i \to j$ with $i \not= j$ are certainly essential
and they do not appear in the Hamiltonian (\ref{C.1e}). Moreover, the
transitions $i \to i$ with a change of occupation number are not correctly
represented in a quantity such as
$$tr \left ( H_1 H_1 \rho_e \right ) = \sum_{nn'} 
<n|H_1|n'><n'|H_1|n> \rho_{en} \  .$$

\noi If the occupied state $|n>$ is considered as the first excited 
state $|i,1>$  of
an oscillator, the contribution of $n'> = |i,2>$ cannot be omitted from the sum
(\ref{C.2e}) if the Hamiltonian (\ref{C.1e}) is used. The interest of a theory
ignoring the constraints of the oscillator model cannot therefore be
disputed.\\

\noi {\large \it The scattering model}\\

Joos and Zeh have proposed a beautiful model of decoherence, when an
object with position $X$ interacts with a bath of particles \cite{9r}.
Decoherence appeared then as an accumulation of scattering effects. The
resulting master equation looked very similar to one occurring in the
oscillator model and this was a very strong hint for some universality
in the mathematical expression of decoherence.  This universal
character has been explained in the present paper, but something would
still be missing if the scattering model were not also derived. This
derivation is non-trivial so that it will be given explicitly.\\

Let us consider for definiteness a spherical solid object with center
$X$. As explained in Section 3, the pressure exerted on it by the outside
gas is included in $H_c$ so that $H_1$ represents the effect of random
collisions of the outside molecules (or photons) on the sphere. Rather
than doing a complete calculation, it will be sufficient to look at one
term in Eq. (\ref{5.4e}), namely (with $\hbar = 1$)
\beq \label{C.3e}
D_1 = - \int_{-\infty}^t dt' \ tr \left \{ H_1 U_c(t-t') \rho_r (t') U_c^{-1} (t-t') \otimes \rho_e H_1^T \right \} \ .
\eeq

One may consid 
er first the case of an environment consisting of
particles having the same momentum, in a pure state $|k>$ : $\rho_e = 
|k><k|$. The
corresponding wave function is a plane wave $\exp (ik.r)$ and, with this
normalization, there is one particle per unit volume. Their flux $\Phi$ is
the velocity $\nu$ of the particle and a sum over one-particle states $|k'>$
amounts to an integration over $dk/(2 \pi )^3$. \\

A few remarks are useful:

1. One can neglect the time evolution factors $U_c(t-t')$, which are slowly
varying. \\

2. Introducing momentum eigenstates $|p>$ of the object and the
outgoing scattering states $|p,k,out>$ associated with the plane 
waves $|p,k> = |p> \otimes |k>$, one can
write
\beq \label{C.4e}
<p_2,k_2,out|H_1|p_1,k_1> = <p_2,k_2|T|p_1,k_1> \delta \left ( p_1 + 
k_1 - p_2 - k_2 \right ) \ ,
\eeq

\noi where $T$ is the (off energy-shell) collision matrix for the scattering
of a particle on the object [45]. \\

3. One can use the invariance of the scattering $T$-matrix under a change
of reference frame. For non-relativistic values of $p$ and taking into
account the large mass of the object (as compared with the particle
mass), a Galilean transformation with small velocity gives
\beq \label{C.5e}
<p_2,k_2|T|p_1,k_1> = <p_2 - p_1, k_2 |T|0,k_1> <p_1,k_1|T|p_2,k_2> = 
<0,k_1|T|p_2 - p_1,k_2> \ . \eeq

4. In view of the delta function in Eq. (\ref{C.4e}), it is enough to 
know $k_1$
and $k_2$ for fixing $p_2 - p_1$ so that the right-hand side of Eq. 
(\ref{C.5e}) can be
written more simply $T(k_1,k_2)$. Conversely, using Fourier transforms, one can
introduce the states $|x,k>$ corresponding to a localized object, whereas out
states $|x,kout>$ involve the same value of $x$ and a scattered 
particle. One thus
gets
\beq \label{C.6e}
<x',k_2out|H_1|x,k_1> = T\left ( k_1,k_2\right ) \delta (x - x') \exp 
\left \{ i \left ( k_1 - k_2 \right ) x \right \} \ .
\eeq

\noi This result has two important consequences: It shows that $H_1$ 
acts like a
function of $X$, although this property shows up only when scattering
states are used. The imaginary exponential in the right-hand side is
moreover typical of the method that was used by Joos and Zeh.\\

5. In view of Eqs. (\ref{5.3e}), and (\ref{C.4e}), one has
\beq \label{C.7e}
<x,k_1|H_1^T|x',k_2, out> = \exp \left \{ i \omega (t - t') \right \} 
\delta (x-x') \exp \left \{ - i\left ( k_1 - k_2 \right ) x \right \} 
\cdot T^*\left ( k_1, k_2 \right ) \ ,
\eeq

where $\omega = E(k_2) - E(k_1)$.\\

6. When throwing out retardation, the integral on $t'$ of the imaginary
exponential in (\ref{C.7e}) gives $\pi \delta (\omega ) - iP(1/ 
\omega )$, where $P$ stands for a Cauchy principal
part. It may be shown however that another term in $D$ originating 
from $H_1^T\rho_1H_1$
cancels the principal part and one must keep only therefore the delta
part.\\

Finally, the matrix element $<x|D_1|x'>$ can be easily computed if one uses the
orthonormal set of outgoing states $\{|k' \ out>\}$ when computing 
the trace $tr$. One
gets
\beq \label{C.8e}
<X|D_1|x'> = \int \left ( dk/(2 \pi )^3 \right ) \pi \delta (\omega ) 
|T(k,k')|^2 \exp \left \{ i(k-k') (x-x')\right \} \rho_r(x, x') \ .
\eeq

\noi But the quantity $dk(2 \pi )^{-3}\delta (\omega )|T(k,k')|^2$ 
has a very simple interpretation: It coincides with
the product $d\sigma \Phi$ of the differential cross-section $d\sigma$ for the
scattering $k \to k'$ times the flux $\Phi$ of the environment particles having
the given momentum $k$. One can then replace the trivial density 
matrix $|k><k|$
by a thermal density and introduce the various different particles in
the gas, thus obtaining:
\beq \label{C.9e}
<x|D_1|x'> = \int \pi d \sigma d \Phi \exp \left \{ i (k-k') (x-x') 
\right \} \rho_r (x, x') \ .
\eeq

Similar results are obtained for the three other terms in $D$ but it will
not be necessary to push the calculation further since, from there on,
it becomes identical with the one by Joos and Zeh. Their method was of
course simpler than the present one, as one expects from an intuitive
approach compared to a technical one. The present calculation shows
however how universal and versatile the fundamental master equation
(\ref{4.6e}) is.

\end{document}